\documentclass[aps,prb,superscriptaddress,longbibliography,floatfix,amsfonts,amssymb,amsmath,11pt,reprint]{revtex4-2}
\usepackage[utf8]{inputenc}
\usepackage{amsmath}
\usepackage{amsfonts}
\usepackage{amssymb}
\usepackage{graphicx}
\usepackage[hidelinks, breaklinks=true,unicode]{hyperref}
\usepackage{ulem} 
\usepackage[product-units=power, separate-uncertainty=true, multi-part-units=single]{siunitx} 
\usepackage[usenames,dvipsnames]{xcolor}
\usepackage{threeparttable}

\DeclareSIUnit\angstrom{\text {\AA}}
\DeclareMathOperator{\sgn}{sgn}
\DeclareUnicodeCharacter{202A}{ }

\newcommand{\dIdU}{$\dd I / \dd V$}
\newcommand{\ddIddU}{$|\dd^2 I / \dd V^2|$}
\newcommand{\dd}{\text{d}}

\newcommand{\etal}{\textit{et al.}}

\newcommand{\JB}[1]{\textcolor{black}{#1}}

\begin{document}

\title{Characterizing the chemical potential disorder in the topological insulator (Bi$_{1-x}$Sb$_x$)$_2$Te$_3$ thin films}

\author{Jens Brede}\email{brede@ph2.uni-koeln.de (he/him/his)}\affiliation{Physics Institute II, University of Cologne, D-50937 K\"oln, Germany}
\author{Mahasweta Bagchi}\affiliation{Physics Institute II, University of Cologne, D-50937 K\"oln, Germany}
\author{Adrian Greichgauer }\affiliation{Physics Institute II, University of Cologne, D-50937 K\"oln, Germany}
\author{‪Anjana Uday}\affiliation{Physics Institute II, University of Cologne, D-50937 K\"oln, Germany}
\author{Andrea Bliesener}\affiliation{Physics Institute II, University of Cologne, D-50937 K\"oln, Germany}
\author{Gertjan Lippertz}\affiliation{Physics Institute II, University of Cologne, D-50937 K\"oln, Germany}
\author{Roozbeh Yazdanpanah}\affiliation{Physics Institute II, University of Cologne, D-50937 K\"oln, Germany}
\author{‪Alexey Taskin}\affiliation{Physics Institute II, University of Cologne, D-50937 K\"oln, Germany}
\author{Yoichi Ando}\email{ando@ph2.uni-koeln.de (he/him/his)}\affiliation{Physics Institute II, University of Cologne, D-50937 K\"oln, Germany}

\date{\today}

\begin{abstract}
We use scanning tunneling microscopy and spectroscopy under ultra-high vacuum and down to 1.7 K to study the local variations of the chemical potential on the surface of  the topological insulator (Bi$_{1-x}$Sb$_x$)$_2$Te$_3$ thin films (thickness 7 -- 30 nm) with varying Sb-concentration $x$, to gain insight into the charge puddles formed in thin films of a compensated topological insulator.
We found that the amplitude of the potential fluctuations, $\Gamma$, is between 5 to 14 meV for quasi-bulk conducting films and about 30 -- 40 meV for bulk-insulating films. The length scale of the fluctuations, $\lambda$, was found to span the range of 13 -- 54 nm, with no clear correlation with $\Gamma$.
Applying a magnetic field normal to the surface, we observe the condensation of the two-dimensional topological surface state into Landau levels and find a weak but positive correlation between $\Gamma$ and the spectral width of the Landau-level peaks, which suggests that quantum smearing from drift motion is the source of the Landau level broadening.
Our systematic measurements give useful guidelines for realizing $(\mathrm{Bi}_{1-x}\mathrm{Sb}_x)_2\mathrm{Te}_3$ thin films with an acceptable level of potential fluctuations.
In particular, we found that $x\approx0.65$ realizes the situation where $\Gamma$ shows a comparatively small value of 14 meV and the Dirac point lies within $\sim$10 meV of the Fermi energy.
\end{abstract}

\maketitle

\section{Introduction}

Inducing superconductivity in a topological insulator (TI) by the proximity effect is a promising way to realize topological superconductivity (TSC)~\cite{Fu2008}. 
In particular, when a bulk-insulating 3D TI is confined to a quasi-one-dimensional nanowire, the topological surface state (TSS) breaks into peculiar subbands, which, when proximitized by a conventional superconductor, can host Majorana zero modes \cite{Cook2011}.
However, while the typical subband spacing in TI nanowires is of the order of meV, the chemical potential at the surface of 3D bulk TIs varies by tens of meV due to disorder \cite{Beidenkopf2011, Chong2020, Okada2012, Fu2013, Lee2015, Pauly2015, Fu2016, Storz2016, Knispel2017}.  
Indeed, recent calculations~\cite{Skinner2013,Huang2021} suggest that typical impurity concentrations of the order of \qty{1e19}{\per\cubic\centi\metre} are already sufficient to obscure any subband signatures in transport properties of TI nanowires.

In 2D TIs, the chemical potential fluctuations due to disorder lead to the formation of charge puddles in the insulating 2D bulk, which influences the transport properties of metallic 1D edge states in many ways:
In HgTe quantum well \cite{Dartiailh2020,Gourmelon2023}, in InAs/InGaSb \cite{Kamata2022}, and in V-doped (Bi$_{1-x}$Sb$_x$)$_2$Te$_3$ \cite{roeper2024}, the charge puddles are believed to be responsible for a reduction of the group velocity of the edge state. 
In another experiment on HgTe quantum well in magnetic fields \cite{Shamim2022}, it was proposed that an accumulation of bulk puddles at the edge of the 2D bulk supports a chiral quantum Hall edge channel coexisting with the helical edge channel and leads to a finite (and opposite) conductance. In the case of quantum anomalous Hall insulators that are realized in ferromagnetic TI films, the charge puddles increase the localization length of the chiral edge state \cite{Zhou2023} or are responsible for the breakdown of the quantum anomalous Hall effect \cite{Lippertz2022}.
It is clear from these studies that precise information on the charge puddles in thin TI materials is indispensable for understanding their transport properties.

Experimentally, scanning tunneling microscopy and spectroscopy (STM/STS), which allows real space mapping of the variations in the electrical potential with sub-nanometer spatial and sub-meV energy resolution, has been employed to study the disorder effects on prototypical TI materials such as $\mathrm{Bi}_2\mathrm{Se}_3$, $\mathrm{Sb}_2\mathrm{Te}_3$, and $\mathrm{Bi}_2\mathrm{Te}_3$.
These STM/STS studies consistently revealed chemical potential fluctuations $\Gamma$ of about 10 -- 60 meV over a typical length scale $\lambda$ of about 10 -- 60 nm at the surface of bulk crystals~\cite{Beidenkopf2011, Chong2020, Okada2012, Fu2013, Lee2015, Pauly2015, Fu2016, Storz2016, Knispel2017} and nanoplatelets~\cite{Parra2017}. However, similar information for thin films is scarce \cite{Scipioni2018}.

In this work, we use STM/STS under ultra-high vacuum (UHV), at 1.7 K and out-of-plane magnetic fields up to 9 T to study the local variations of the chemical potential at the surface of (Bi$_{1-x}$Sb$_x$)$_2$Te$_3$ thin films grown on sapphire [Al$_2$O$_3$(0001)]. To see the possible effect of screening due to a superconductor at the bottom of the film, we also measured (Bi$_{1-x}$Sb$_x$)$_2$Te$_3$ films on superconducting Nb prepared by the ``flip-chip" (FC) technique \cite{Floetotto2018}. Although our samples are clean enough to present clearly resolved Landau-level peaks in perpendicular magnetic fields, the observed potential fluctuations of up to $\sim$40 meV indicate that the effect of Coulomb disorder is relatively strong in (Bi$_{1-x}$Sb$_x$)$_2$Te$_3$ thin films, particularly when the film is bulk-insulating. Nevertheless, at $x \approx$ 0.65, one can obtain a quasi-bulk-insulating sample with a relatively low potential disorder of 14 meV.

\section{Experimental Methods}

(Bi$_{1-x}$Sb$_x$)$_2$Te$_3$ films were grown with the molecular beam epitaxy (MBE) technique as described previously \cite{Yang2014, Taskin2017}. After cooling to 310 K, samples were capped by 10 -- 20 nm of Te to protect the films during transfer. The Te-capped samples were removed from the MBE chamber and transferred ex-situ to the STM system. Typical transfer times are less than 5 minutes, minimizing the exposure of the capped films to ambient conditions.
Inside the STM preparation chamber (pressure $p < 2\times 10^{-10}$~mbar), the samples are first outgassed at about 400~K for about 30~min, subsequently heated up to about 540~K (in about 10~min), kept at 540~K for $< 5$~min, and then cooled to room temperature. The pressure in the chamber before turning off the heater is typically well below $1\times 10^{-9}$~mbar, such that the Te is completely evaporated.
Some films were transferred from the sapphire substrate onto a superconducting Nb film using a flip-chip technique \cite{Floetotto2018}: First, an amorphous Nb layer of several tens of nm was deposited onto the (Bi$_{1-x}$Sb$_x$)$_2$Te$_3$ surface, and the top Nb surface is subsequently glued to a metallic substate to flip the film; the sample is then brought into the STM preparation chamber and the sapphire substrate is removed in UHV by cleaving, which usually occurs between the (Bi$_{1-x}$Sb$_x$)$_2$Te$_3$ and sapphire rather than between (Bi$_{1-x}$Sb$_x$)$_2$Te$_3$ and Nb due to the sticky nature of Nb. One can thus obtain a clean (Bi$_{1-x}$Sb$_x$)$_2$Te$_3$ surface suitable for STM observations.

After obtaining a clean surface in the STM preparation chamber either by Te evaporation or by cleaving, the samples are transferred in-situ into the STM main chamber and cooled down.
STM experiments are carried out under UHV conditions with a commercial system (Unisoku USM1300) operating at a base temperature of $1.7$~K unless stated otherwise. Topography and \dIdU{} maps are recorded in the constant-current mode. Spectroscopy data is obtained by first stabilizing for a given setpoint condition and then disabling the feedback loop. \dIdU{} spectra are then recorded using a lock-in amplifier by adding a small modulation voltage $V_{\text{mod}}$ with frequency $F_\mathrm{mod}=311$~Hz to the sample bias voltage $V$. 
Tips have been prepared by Ar ion sputtering (at an argon pressure of $3\times10^{-6}$~mbar and a voltage of $1$~kV), followed by repeated heating by electron bombardment ($\approx$ $40$~W) for $<10$~s. Further tip forming is done by scanning on the Cu(111) surface until a clean signature of the surface state is obtained in spectroscopy. Data are processed using Igor Pro.

\section{Basic characterizations}

\begin{figure}[h]
\includegraphics[scale=.8]{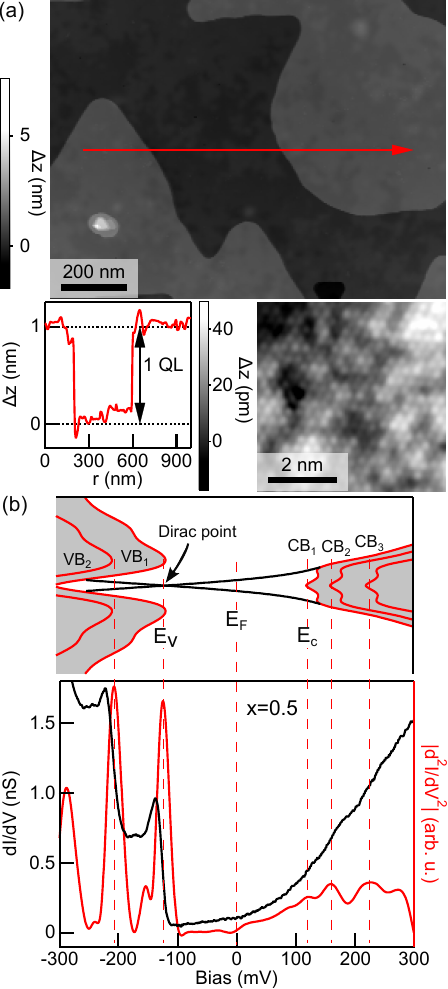}
\caption{\textbf{Basic characterization of (Bi$_{1-x}$Sb$_x$)$_2$Te$_3$ films with STM/STS.} (a) The large topographic image shows extended atomically-flat terraces having a quintuple layer (QL) step height ($\sim$1 nm). The height profile was taken along the red arrow. The atomic resolution image shows the hexagonal lattice of the top Te-layer. 
(b) Example STS data for $x$ = 0.5. The \dIdU{} spectrum (black trace in the bottom panel) has a U-shape around $E_\mathrm{F}$ (= zero Bias). We phenomenologically define the onsets of the bulk valence band $E_\mathrm{v}$ and bulk conduction band $E_\mathrm{c}$ from the maxima in the calculated $|\dd^2I/\dd V^2|$ data (red trace) as indicated. The band structure of (Bi$_{1-x}$Sb$_x$)$_2$Te$_3$ is sketched for illustration above the STS data. For this sample, the Dirac point coincides with $E_\mathrm{v}$. Setpoints: (a) $I_0=20$~pA, $V_0=900$~mV; atomic resolution: $I_0=0.5$~nA, $V_0=100$~mV; (b) $I_0=0.2$~nA, $V_0=300$~mV, $V_\mathrm{mod}=5$~mV$_\mathrm{p}$.}
\label{fig:STM_1}
\end{figure}

A typical STM topography of the surface of our (Bi$_{1-x}$Sb$_x$)$_2$Te$_3$ film is shown in Fig.~\ref{fig:STM_1}(a), which shows that the film has quintuple layer (QL) step heights of about 1~nm (line profile taken along the red arrows).  
A low surface roughness of this film is evident with only two terrace heights being visible in the entire field of view.
The atomic resolution imaging shows the characteristic hexagonal lattice of the top Te-layer with a lattice spacing of about \qty{4.3}{\angstrom}, that is consistent with the (Bi$_{1-x}$Sb$_x$)$_2$Te$_3$ structure. 

A typical \dIdU spectrum, which is roughly proportional to the local density of states (LDoS) of the sample, is shown for $x$ = 0.5 in Fig.~\ref{fig:STM_1}(b):
The Fermi energy $E_\mathrm{F}$ lies in the U-shaped minimum of the bulk band gap and the energy position of the bulk valance band top $E_\mathrm{v}$ is easily recognizable as a sharp step-like increase in the \dIdU{} spectrum at $E_\mathrm{v} \approx \qty{-120}{\meV}$. In contrast, the energy position of the bulk conduction band bottom $E_\mathrm{c}$ is more difficult to identify, as it causes a more subtle increase at $E_\text{c} \approx \qty{120}{\meV}$. Nevertheless, by numerically calculating the derivative of the differential tunnel conductance $\dd^2I/\dd V^2$ and phenomenologically defining $E_\mathrm{c}$ as the maxima in $|\dd^2I/\dd V^2|$, we can objectively determine $E_\mathrm{c} \approx$ 120 meV. The same definition gives correct $E_\mathrm{v} \approx -120$ meV for the bulk valence band.
In this film, the Dirac point $E_\mathrm{D}$ lies almost exactly at the bulk valence band top as determined by Landau level spectroscopy discussed in Sec.~\ref{sec:LandauLevelandEDP} and the supplement~\cite{SM}. We will discuss the properties of the TSS in Sec.~\ref{sec:LandauLevelandEDP}.
The additional maxima in $|\dd^2I/\dd V^2|$ at lower and higher energies than $E_\mathrm{v}$ and $E_\mathrm{c}$ present a spacing of around 90 meV. They are attributed to the quantization of the bulk bands caused by the finite film thickness, as discussed in Appendix.~\ref{Sec:QWS}. 

\section{Amplitude and length scale of potential fluctuations}
\label{sec:disorder}

\begin{figure}[h]
\includegraphics[scale=.8]{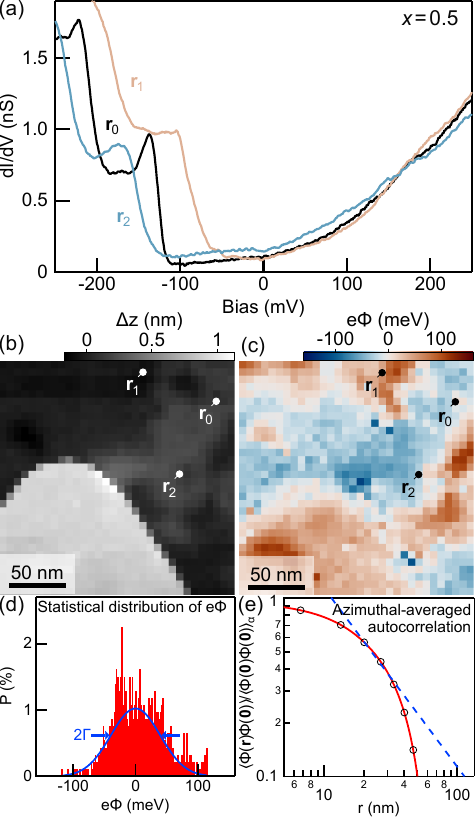}
\caption{\textbf{Chemical potential fluctuations at the surface.} (a) \dIdU{} spectra of a $x$ = 0.5 sample recorded at ${\bf r}_{0}$, ${\bf r}_{1}$, and ${\bf r}_{2}$ shown in (b); the data at ${\bf r}_{0}$ is the one shown in Fig.~\ref{fig:STM_1}(b). (b) Topography of the area where the data in (a) were measured. (c) The spatial distribution of $e\phi$ presenting puddles. (d) Fitting the histogram of $e\phi$ with a Gaussian distribution (blue trace) gives the amplitude of the chemical potential fluctuations $\Gamma$ as the standard deviation. (e) Average puddle size $\lambda$ is defined as the distance $r$ where the decay of the azimuthal mean of the autocorrelation of the data in (c), depicted with a red curve, becomes faster than a $1/r$ decay (dashed blue line); in this case, $\lambda \approx$ 30 nm. 
\JB{Excluding the large terrace at lower left in the view does not change $\lambda$.} 
Setpoints: (a,b): $I_0=0.2$~nA, $V_0=300$~mV, $V_\mathrm{mod}=5$~mV$_\mathrm{p}$.}
\label{fig:Potential}
\end{figure}

We employ the well-established rigid shift model~\cite{Morgenstern2003, Beidenkopf2011, Chong2020} to measure the local potential disorder at the surface given as $e\phi=E_\mathrm{v}-\langle E_\mathrm{v} \rangle=E_\mathrm{c}-\langle E_\mathrm{c}\rangle$, where $\langle ... \rangle$ denotes the statistical mean. To this end, we record STS grids covering areas of $\qtyrange{2500}{62500}{\nm^2}$ with resolutions of \qtyrange{16}{100}{\nm^2}, and determine $E_\mathrm{v}$ and $E_\mathrm{c}$ for each spectrum and calculate $e\phi$ at each location of the grid. As shown in Appendix \ref{sec:TechnicalDetailsDisorderPotentialCharacterization}, we have actually confirmed the rigid-shift nature of the position-dependent spectra.
The topography simultaneously acquired with one such STS grid is shown in Fig.~\ref{fig:Potential}(b) and spectra from the grid at ${\bf r}_{0}, {\bf r}_{1}, {\bf r}_{2} $ are plotted in Fig.~\ref{fig:Potential}(a). 
The spectrum taken at ${\bf r}_{0}$ is the one plotted in Fig.~\ref{fig:STM_1}(b) and serves as a reference to illustrate the rigid shift of the characteristic features at $E_\mathrm{v}$ and $E_\mathrm{c}$ in the spectra taken at ${\bf r}_{1}$ and ${\bf r}_{2}$. 


Plotting $e\phi$ as a function of position in Fig.~\ref{fig:Potential}(c), one can see the spatial distribution of the potential disorder, where the extended excess-electron puddles with negative $e\phi$ (blue) and fewer-electron puddles with positive $e\phi$ (red) spanning tens of nanometer are conveniently illustrated. Note that these puddles are chemical potential fluctuations in a metallic background and they are different from electron and hole puddles formed in an insulating background \cite{Skinner2013a}, although both are caused by Coulomb impurities.
We define the amplitude of the chemical potential fluctuations $\Gamma$ as the width of the Gaussian distribution fit to the histogram of $e\phi$ as shown in Fig.~\ref{fig:Potential}(d) \cite{Skinner2013a, Huang2021a, Boemerich2017}, and the typical length scale of the puddle $\lambda$ is defined as the distance where the azimuthal average of the autocorrelation in the map of $e\phi$ starts to decay more rapidly than $\sim 1/r$, see Fig.~\ref{fig:Potential}(e). In the example shown here, $\Gamma\approx$ 40 meV and $\lambda \approx$ 30 nm.
The choice of a faster-than-$1/r$ decay for the definition of $\lambda$ is motivated by the intuition that it marks the distance at which the Thomas-Fermi polarization bubble  screens the charged impurities. 
The $\lambda$ values defined this way are consistent with what one can infer in Fig.~\ref{fig:Potential}(c) as the puddle size (additional examples are given in the supplement).

\begin{figure*}[t]
\includegraphics[scale=.85]{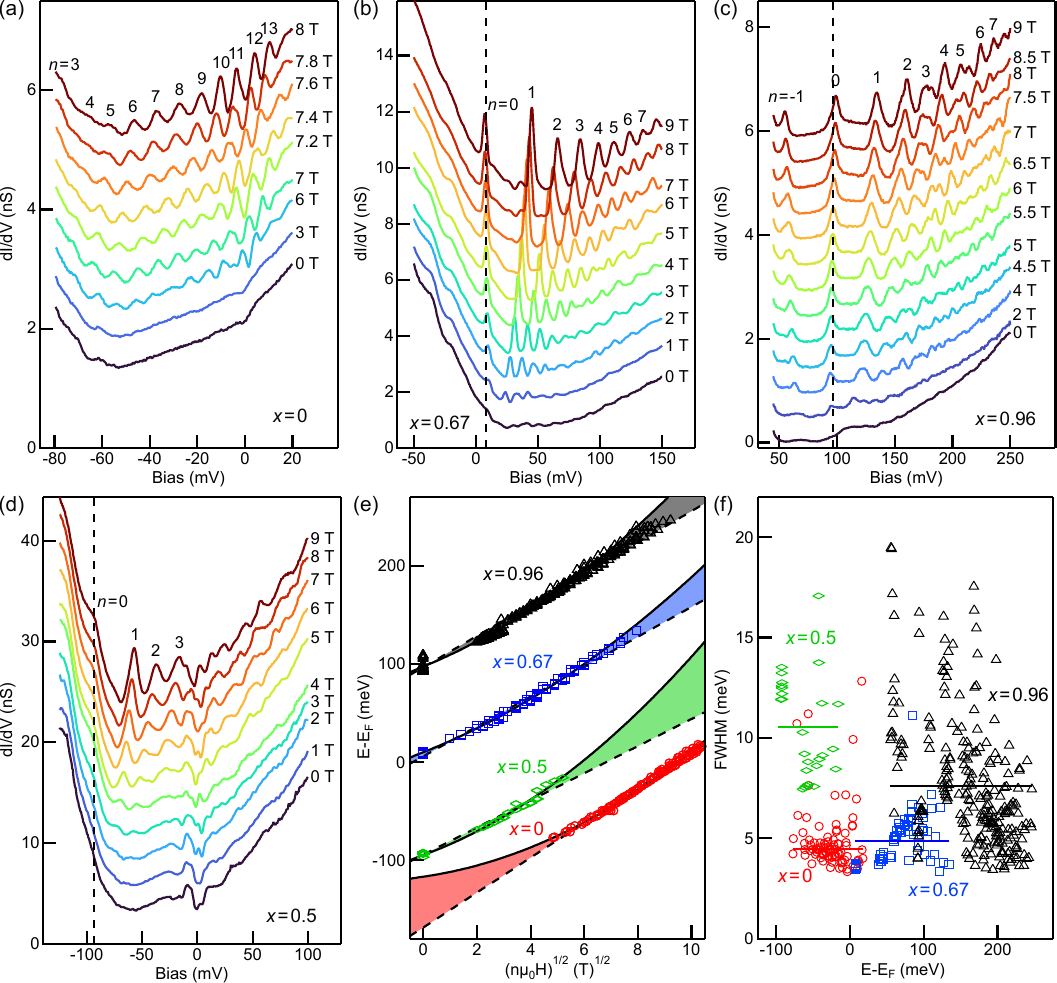}
\caption{\textbf{Landau level spectra.} (a-d) Spectra of the films with different $x$ values on sapphire (a,b,c) and Nb (d). The spectra in each panel have been vertically shifted for clarity and the Landau level index $n$ is given for the highest magnetic field data. (e) Eigenenergies of the Landau levels were collected from the data in (a-d). The fitting of the data using Eq.~\ref{eq:E_n} is shown for two cases: neglecting the $H$-linear term (dashed lines) and with the full formula (solid lines). Shaded regions mark the area between the two fits. (f) FWHM of the Landau-level peaks analyzed after background subtraction. The horizontal lines give the arithmetic mean $\langle \mathrm{FWHM} \rangle$ for a given $x$. Setpoints: (a) $I_0=50$~pA, $V_0=20$~mV;
(b) $I_0=0.2$~nA, $V_0=150$~mV; (c) $I_0=0.2$~nA, $V_0=250$~mV; (d) $I_0=-0.5$~nA, $V_0=-100$~mV; (a,b,c,d) $V_\mathrm{mod}=2$~mV$_\mathrm{p}$.}
\label{fig:LandauLevel}
\end{figure*}

\section{Landau levels and Dirac point}
\label{sec:LandauLevelandEDP}

The LDoS of an idealistic massless two-dimensional Dirac cone of the TSS increases linearly with $E-E_\text{D}$ and it is in principle straightforward to determine the Dirac point energy $E_\text{D}$ from the linear slope of the \dIdU{} signal inside the bulk band gap. In practice, however, complications due to a finite curvature~\cite{Fu2014a, Fu2016} and broadening of the bulk valence band top by shallow acceptor states lead to large uncertainties in $E_\text{D}$ determined in such a way, especially for the $x$ range where $E_\text{D} \lesssim E_\mathrm{v}$. 
On the other hand, Landau level spectroscopy~\cite{Morgenstern2003, Hashimoto2008, Cheng2010, Jiang2012a, Hanaguri2010, Pauly2015, Storz2018, Fu2014a, Fu2016, Bagchi2022} is ideally suited to isolate the spectral features of the TSS by looking at the response of the LDoS to an external magnetic field $\mu_0 H$ applied normal to the surface.

In the case of the TSS, the eigenenergy $E_n$ of the $n$-th Landau level is approximately given as
\begin{equation}
E_n \approx E_\mathrm{D}+\sgn{(n)}v_\mathrm{D} \sqrt{2\hbar e|n|\mu_0H}+ \frac{\hbar e}{m_\mathrm{eff}}n\mu_0H,
\label{eq:E_n}
\end{equation}
with $E_\mathrm{D}$ the Dirac point energy and $v_\mathrm{D}$ the Fermi velocity. The last term in Eq.~\ref{eq:E_n} is a correction due to a finite curvature of the Dirac cone~\cite{Cheng2010, Jiang2012a, Hanaguri2010, Pauly2015, Storz2018, Fu2014a, Bagchi2022} giving rise to the effective mass $m_\mathrm{eff}$. 
For simplicity, corrections due to the Stark-effect~\cite{Okada2012, Fu2014a} and additional effective Zeeman effect~\cite{HernangomezPerez2013, Fu2016} are disregarded. 

Before measuring the magnetic-field dependence of the spectra discussed below, we typically performed a rough spatial sampling of several ($<10$) positions {\bf r} in the highest magnetic field, and performed the field dependence measurements at the {\bf r} position where the sharpest spectrum was obtained in the sampling. As will be discussed in Sec.~\ref{sec:discussions}, these regions likely correspond to local extrema in the fluctuating potential.

In Fig.~\ref{fig:LandauLevel}(a)-(d) we show spectra for $x$ = 0, 0.5, 0.67, and 0.96 in various magnetic fields, where well-defined peaks in the LDoS appear with increasing magnetic field.
For each $x$ value, we indicate the Landau level index $n$ on the spectra taken at the highest field, and the vertical dashed line marks the position of $E_\text{D}$ when it is in the measured bias range.
In the case of $x$ = 0 (i.e. Bi$_2$Te$_3$) shown in Fig.~\ref{fig:LandauLevel}(a), the first Landau level resolved has the index $n=3$ and lies about 80 meV below the Fermi energy $E_\text{F}$; Landau levels with lower indices are completely masked by the LDoS stemming from the bulk valence band and only Landau levels with $n \geq 6$ are seen in the bulk band gap.
On the other hand, for $x=0.96$ shown in Fig.~\ref{fig:LandauLevel}(c), even the first hole-like Landau level ($n=-1$) is seen in the bulk-band gap and it is located at $\sim$50 meV above $E_\text{F}$.
In agreement with a previous study~\cite{Scipioni2018}, we find that for $x=0.67$ [Fig.~\ref{fig:LandauLevel}(b)], $E_\text{D}$ lies within $\sim$10 meV of $E_\text{F}$ and, although $E_\text{D}$ is slightly below $E_\mathrm{v}$, the zeroth Landau level is still not smeared by bulk carriers. 

Note that the zeroth Landau level in Fig.~\ref{fig:LandauLevel}(b) slightly shifts towards lower energy with increasing field, while it shifts in the opposite direction in Fig.~\ref{fig:LandauLevel}(c). This magnetic-field dependence of the zeroth Landau level is not captured in Eq.~\ref{eq:E_n} but is understood as an effective Zeeman effect $\delta E$ that is proportional to the gradient of the local potential $e\phi({\bf r})$ and the magnetic length $l_\mathrm{H}=\sqrt{\hbar/(e|\mu_0H|)}$ squared, namely, $\delta E \propto \nabla_{\bf{r}}e\phi({\bf r}) l_\mathrm{H}^2$ \cite{HernangomezPerez2013, Fu2016}. Therefore, the opposite direction of the shift in Figs.~\ref{fig:LandauLevel}(b) and \ref{fig:LandauLevel}(c) suggests that the two sets of spectra should have been taken near a potential minimum and maximum, respectively.

\JB{Another effect not captured by Eq.~\ref{eq:E_n} is the lifting of the degeneracy of Landau subbands with different total angular momentum ($j_z$). In the first approximation, as shown by Fu \etal~\cite{Fu2014a}, Landau orbits with higher $j_z$ drift at larger $| {\bf r }|$ and experience a larger potential $e\phi({\bf r})$, which lifts the degeneracy. This effect is likely responsible for the splitting of Landau levels with $n=3,5$ in Fig.~\ref{fig:LandauLevel}(c). Additional spectra showing the splitting of Landau levels can be found in Fig.~S10 of the supplement~\cite{SM}.}

As for the dependence of the Landau levels on $x$, we note that for $x=0.5$, the zeroth Landau level to identify $E_\text{D}$ is sometimes visible [as in Fig.~\ref{fig:LandauLevel}(d)] but it can also be completely masked by the bulk valence band (we measured four samples with $x$ = 0.5, see Table \ref{tab:overviewsamples}).
In the latter case, we determined $E_\text{D}$ from Eq.~\ref{eq:E_n} as shown in Fig.~\ref{fig:LandauLevel}(e), where one can see that, even by neglecting the last term in Eq.~\ref{eq:E_n}, the experimental data can be reasonably fit with $v_\mathrm{D}=\qtyrange{3.8e5}{6.2e5}{\meter\per\second}$, which is in good agreement with previous studies~\cite{Okada2012, Pauly2015, Storz2016}. Slight improvements are achieved by using the full expression (solid lines) with $m_\mathrm{eff}=\qtyrange{0.1}{0.5}{m_e}$ and $v_\mathrm{D}=\qtyrange{2.8e5}{5e5}{\meter\per\second}$ ($m_e$ is the free electron mass).

\begin{figure}[htb]
\includegraphics[scale=0.9]{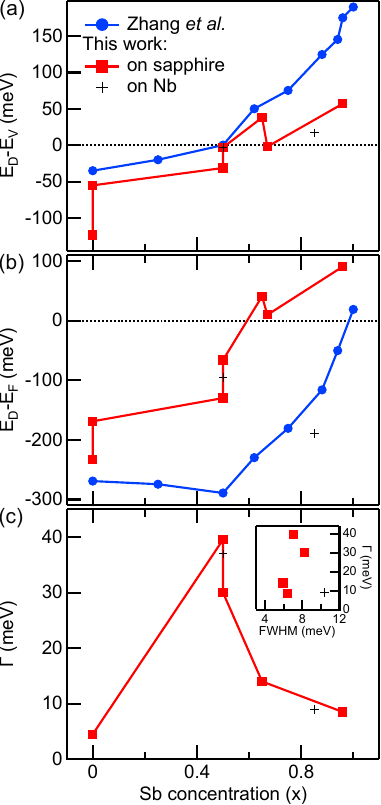}
\caption{\textbf{Evolution of the position of the Dirac point and the amplitude of the potential fluctuations as a function of $x$.} (a) $E_\mathrm{D}-E_{\rm v}$ vs $x$ plotted for our films on sapphire and Nb, along with the data extracted from the report by Zhang {\it et al.} \cite{Zhang2011}. (b) Similar comparison for $E_\mathrm{D}-E_{\rm F}$.  (c) $x$-dependence of $\Gamma$. In the inset, obtained $\Gamma$ is plotted against $\langle \mathrm{FWHM}\rangle$ measured on the same sample, which was possible for 5 samples (see Table \ref{tab:overviewsamples}). The data from the $x=0.86$ on-Nb sample in this inset shows an unusually large $\langle \mathrm{FWHM}\rangle$ for a small $\Gamma$, which is possibly due to a large number of Te-vacancies in this $n$-type sample ($E_{\rm c} < 0$); note that all other samples had $E_{\rm c} > 0$ (see Table I). Except for this data point, the data in the inset points to a weak but positive correlation between $\Gamma$ and $\langle \mathrm{FWHM}\rangle$.}
\label{fig:EDvsX}
\end{figure}

To analyze the line-shape, we extract the full-width at half maximum (FWHM) of each Landau level peak after subtracting a smooth background (details on the background subtraction and fitting are found in Appendix~\ref{sec:LandauLevelDisorder}), and the results are collected in  Fig.~\ref{fig:LandauLevel}(f).
The FWHM of the $x=0$ and $x=0.67$ samples are essentially in the 3--8 meV range, which agrees well with previous reports on bulk crystals~\cite{Hanaguri2010,Scipioni2018,Storz2018}. Moreover, as originally reported by Hanaguri \etal~\cite{Hanaguri2010} for bulk $\mathrm{Bi}_2\mathrm{Se}_3$, the Landau levels in these samples sharpen  near $E_\text{F}$. In particular, for $x=0.67$, the FWHM of the $n=0$ peak which is located near $E_{\rm F}$ is only $\sim$4 meV. Intriguingly, the data from the $x=0.96$ sample [black symbols in Fig.~\ref{fig:LandauLevel}(f)] show a different trend: The FWHM tends to be reduced when the peaks are further away from $E_\mathrm{F}$. A similar trend was observed in bulk $\mathrm{Sb}_2\mathrm{Te}_3${} by Storz \etal~\cite{Storz2018}, while an opposite trend, an increase in FWHM with $E-E_\mathrm{F}$, was reported for $\mathrm{Sb}_2\mathrm{Te}_3$ thin-films by Jiang \etal~\cite{Jiang2012a} and for bulk $\mathrm{Sb}_2\mathrm{Te}_3$ by Pauly \etal~\cite{Pauly2015}. The origin of the complex evolution of the FWHM with energy is beyond the scope of this paper. 

\begingroup
\begin{table*}
	\begin{tabular}{l c c c c c c c  c c c}
		\hline\hline
        Sample & x & t & N$_\mathrm{d}$ & $E_\mathrm{v}$ & $E_\mathrm{c}$ & $E_\mathrm{D}$ & $v_\mathrm{D}$ & $\langle \mathrm{FWHM}\rangle$ & $\Gamma$ & $\lambda$ \\ 
        ~ & ~ & (\unit{\nm}) & (\qty{1e19}{\per\cubic\cm}) & (meV) & (meV) & (meV) & (\qty{1e5}{\meter\per\second})&  (meV) &  (meV) & (\unit{\nm}) \\ \hline
        MBE4 2022 Apr07 & 0 & 17 &   1.2 & -110 & 31 & -233 & $4.6$ & ~ & 5 & $\sim 13$ \\ 
        MBE4 2022 Apr28B & 0 & 12 &   ~ & -114 & 47 & -169 & 4.9 & 4.4 & ~ & ~ \\ 
        MBE3 2022 Oct28D2 & 0.5 & 9 &  ~ & -127 & 121 & -130 & 5.2 & 7 & 40 & $\sim 30$\\
        MBE3 2022 Oct28D2 FC$^*$ & 0.5 & 7 &  ~ & -120 & 185 & ~ & ~ & ~ & 37 & $\sim 54$ \\
        MBE3 2022 Oct28C2 & 0.5 & 9 &  ~ & -35 & 212 & -66 & 6.2 & 8.2& 30 & $\sim 32$ \\
        MBE3 2022 Jun6C FC$^*$ &  0.5 & 12 &  ~ & -92 & 116 & -95 & 3.8 & 9.4 & ~ & ~ \\        
        MBE2 2020 Oct9B & 0.65 & 30 &  ~ & 3 & 240 & 41 & 4.6 & 5.9 & 14 & $\sim 20$ \\
        MBE3 2022 Jun5B2 & 0.67 & 19 &  29 & 12 & 207 & 10 & 4.2 & 5.9 & ~ & ~ \\
        MBE1 2021 Sep8 FC$^*$ & 0.86 & 7 &  ~ & -206 & -5 & -189 & 5.1 & 10.4& 9 & ~ \\
        MBE3 2022 Apr13A & 0.96 & 13 & 50 & 34 & $273$ & 91 & 4.4 & 6.4& 9 & $\sim 45$ \\ 
		\hline\hline
	\end{tabular}
	\caption{{\bf Summary of the samples and STM results.} Sb concentration $x$, film thickness $t$, defect density $N_\mathrm{d}$, bulk valence band top $E_\mathrm{v}$, bulk conduction band bottom $E_\mathrm{c}$, Dirac point $E_\mathrm{D}$, Dirac velocity $v_\mathrm{D}$, average Landau level width $\langle \mathrm{FWHM}\rangle$, potential disorder amplitude $\Gamma$, and puddle size $\lambda$ are indicated for all the samples measured in this study. The films marked $^*$ are flip-chip samples on Nb. The $E_\mathrm{D}$ and the $v_\mathrm{D}$ of sample MBE4 2022 Apr07 were determined from quasi-particle interference~\cite{SM}. $E_\mathrm{c}$ of samples MBE2 2020 Oct9B and MBE3 2022 Apr13A were estimated by linear extrapolation~\cite{SM}.}
	\label{tab:overviewsamples}
\end{table*}
\endgroup

\section{$x$-dependence of the TSS and $\Gamma$}
\label{sec:comparison}

We have performed the analysis described in Sec. \ref{sec:disorder} and Sec. \ref{sec:LandauLevelandEDP} for many {$(\mathrm{Bi}_{1-x}\mathrm{Sb}_x)_2\mathrm{Te}_3$} films with the thickness of 7 -- 30 nm and $x$ values from 0 to 0.96, and the results are summarized in Table~\ref{tab:overviewsamples}.
One can immediately see that with increasing $x$, the Dirac point energy $E_{\rm D}$ shifts up in relation to both $E_{\rm v}$ and $E_{\rm F}$, as plotted in Figs. \ref{fig:EDvsX}(a) and \ref{fig:EDvsX}(b). In agreement with the photoemission work \cite{Zhang2011}, we find that the Dirac point is above the bulk valence band top for $x\gtrsim 0.5$. However, as discussed above, in our Landau-level spectroscopy we find the Dirac point is only consistently well-visible for $x\gtrsim 0.65$ and frequently obscured in $x \approx 0.5$ films due to the overlap with the bulk valance band. Also, in agreement with the previous STM work \cite{Scipioni2018}, we find the Dirac point is the closest to the Fermi level for $x = 0.67$.
Interestingly, our data of $E_{\rm F}$ are systematically shifted compared to the photoemission data \cite{Zhang2011}, which may be attributed to surface band bending~\cite{Cheng2010, Frantzeskakis2017, Bagchi2022} or to higher growth and post-annealing temperatures which were reported to lead to more $n$-type films~\cite{Scipioni2018}.

Next, let us look at the $x$-dependence of the potential disorder amplitude $\Gamma$ shown in Fig. \ref{fig:EDvsX}(c). Interestingly, $\Gamma$ peaks at $x \approx 0.5$, at which $E_{\rm F}$ is located roughly at the middle of the $\sim$200 meV bulk band gap. This suggests that reduced screening by bulk carriers enhances the strength of the disorder effect. In this regard, our samples can be roughly separated into two groups, bulk-insulating films and quasi-bulk-conducting films depending on the position of $E_{\rm F}$, where the former shows $\Gamma \approx$ 30 -- 40 meV while the latter shows $\Gamma \approx$ 5 -- 14 meV. Quantitatively speaking, the films fall into the latter category when $E_{\rm F}$ is within 30 -- 50 meV from a bulk band edge. There is no clear correlation between $\Gamma$ and $\lambda$.

Our data on the flip-chip films on Nb, whose data are also plotted in Fig. \ref{fig:EDvsX}, show that the measured potential disorder amplitude is essentially unaffected by the superconducting Nb film of about 60 nm thickness, suggesting that the additional screening provided by the superconductor at the bottom surface leaves the top surface probed by STM largely unaffected. This is true for both the bulk-conducting film with $\Gamma \approx$ 9 meV and the bulk-insulating film with $\Gamma \approx$ 37 meV on Nb. 


\begingroup
\begin{table*}
	\begin{tabular}{c c c c c c c}
		\hline\hline
		Reference  &  Material & thickness & $ 4\Gamma$ & $\lambda$ & $v_\mathrm{D} $& T\\
		& & (\unit{\nm}) & (meV)  & (\unit{\nm}) & ($\qty{1e5}{\meter\per\second}$)& (\unit{\kelvin})\\ \hline
		Beidenkopf \etal~\cite{Beidenkopf2011} &  $\mathrm{Bi}_{2-x}\mathrm{Mn}_x\mathrm{Te}_3$ & $\infty$ & 40 &  24 & 2 & 4\\
		& $\mathrm{Bi}_{2-x}\mathrm{Ca}_x\mathrm{Te}_3$& $\infty$ & 20 & & & 4\\
		& $\mathrm{Bi}_{2-x}\mathrm{Mn}_x\mathrm{Se}_3$& $\infty$ & 20 & & & 4\\
        Dai \etal~\cite{Dai2016} & $\mathrm{Bi}_2\mathrm{Se}_3$ & $\infty$ & $50^{**}$ & 20 to 30$^{**}$&  & $4.5$ \\
		Okada \etal~\cite{Okada2012} & $\mathrm{Bi}_2\mathrm{Te}_3${} & $\infty$ & 12 & 50 & $4.7\pm0.6$ & 4 \\
		Lee \etal~\cite{Lee2015,Chong2020} & $\mathrm{Cr}_{x}\left(\mathrm{Bi}_{0.1}\mathrm{Sb}_{0.9}\right)_{2-x}\mathrm{Te}_3$ & $\infty$ & 30 & 20 & 5 & 4.5\\
		Chong \etal~\cite{Chong2020,Chong2020a} & $\mathrm{Cr}_{0.08}\left(\mathrm{Bi}_{0.1}\mathrm{Sb}_{0.9}\right)_{1.92}\mathrm{Te}_3$ & $\infty$ & 15& 50 & & 0.3 \\
		& $\left(\mathrm{Bi}_{0.1}\mathrm{Sb}_{0.9}\right)_{2}\mathrm{Te}_3$ & $\infty$& \numrange{15}{40} & \numrange{20}{50} & \numrange{3}{4} & 0.3 \\
		Pauly \etal~\cite{Pauly2015} & $\mathrm{Sb}_2\mathrm{Te}_3${} & $\infty$ & 40 & & $4.9 \pm 0.2$  & 6\\
		Fu \etal~\cite{Fu2014a,Fu2016,Fu2016} & $\mathrm{Bi}_2\mathrm{Te}_2\mathrm{Se}$& $\infty$ & \numrange{10}{30} & \numrange{20}{30} & & \\
			& $\mathrm{Bi}_2\mathrm{Se}_3${}& $\infty$ & \numrange{30}{50}  & \numrange{50}{60} &  & 1.5 \\
		Storz \etal~\cite{Storz2016} & $\mathrm{Sb}_2\mathrm{Te}_3$ & $\infty$ & 15 & 40 & \numrange{4.6}{5.2} & 4\\		
		Knispel \etal~\cite{Knispel2017} & $\mathrm{Bi}\mathrm{Sb}\mathrm{Te}\mathrm{Se}_2$& $\infty$ & \numrange{30}{60} & 40-50 & & \numrange{5.5}{77}  \\
		Parra \etal~\cite{Parra2017} & {$\mathrm{Bi}_2\mathrm{Te}_3$}& 8 to 30 & 2 to 30 & 6 to 15 & & 78 \\
		this work & (Bi$_{1-x}$Sb$_x$)$_2$Te$_3$ & \numrange{7}{30} & \numrange{20}{160}   & \numrange{13}{54}  & \numrange{3.6}{6.2} & \numrange{0.35}{1.7}\\	
		\hline\hline
	\end{tabular}
	\caption{{\bf Potential disorder amplitude $\Gamma$ and puddles size $\lambda$ reported in the literature for TIs.} The collected data were measured at the surface with STM on bulk crystals (indicated with thickness $\infty$) or thin films. In publications where no statistical analysis was performed and only numbers for the maximum potential variations observed were provided, we assume these maxima to be an estimate of $4\Gamma$ (corresponding to $95\%$ confidence interval) of a Gaussian distribution. $^{**}$Crystals with optimized growth conditions were reported to have less disorder.
	}
	\label{tab:puddles}
\end{table*}
\endgroup

\section{Discussions and Conclusion}\label{sec:discussions}

So far we have shown the existence of relatively large potential fluctuations in all of our samples. Now let us briefly discuss its origin. In (Bi$_{1-x}$Sb$_x$)$_2$Te$_3$, the carrier density is controlled by compensation doping, where electron (hole) carriers in the bulk are minimized by countering them with compensating acceptors (donors). These acceptors and donors cause a random distribution of Coulomb impurities, leading to random potential fluctuations and creating charge puddles~\cite{Huang2021a}.
For a TI film of thickness $t$, dielectric constant $\epsilon\approx 100$~\cite{Knispel2017}, Dirac velocity $v_\mathrm{D}\approx\qty{4.8e5}{\meter\per\second}$, the Dirac point as well as the Fermi energy in the middle of the bulk band gap, and in the limit of strong disorder, the amplitude of potential fluctuations $\Gamma$ and the puddle size $\lambda$ are determined by the density of Coulomb impurities $N_\mathrm{d}$ and the effective fine structure constant $\alpha_\mathrm{e} \equiv \alpha \frac{c}{\epsilon v_\mathrm{D}}\approx$ 0.0456 in the following way \cite{Huang2021a}:
\begin{align}
\Gamma = \left(\frac{\pi^3}{2}\right)^{1/6}\frac{e^2N_\mathrm{d}^{1/3}}{4\pi\epsilon_0\epsilon\alpha_\mathrm{e}^{2/3}},\\
\lambda = 2^{-7/6}\frac{\sqrt{t}}{\alpha_\mathrm{e}^{2/3}N_\mathrm{d}^{1/6}},
\end{align}
with $\alpha=1/137$, $c$ the speed of light, and $\epsilon_0$ the vacuum permittivity.
As explained in Appendix \ref{Sec:Defects}, we have estimated the defect density $N_\mathrm{d}$ of $\qtyrange{1.2e19}{5e20}{\per\cubic\cm}$ in our films, which is in agreement with the literature~\cite{Jiang2012a,Pauly2015,Knispel2017}. These $N_\mathrm{d}$ values correspond to $\Gamma$ of 41 -- 140 meV and $\lambda$ of 12 -- 23 nm for $t$ =10 nm, which are consistent with our experimental data. Note that the calculated $\Gamma$ ($\lambda$) should be seen as an upper (lower) bound since our $N_\mathrm{d}$ overestimates the amount of Coloumb impurities when not every defect acts as a charge dopant~\cite{Dai2016}. Moreover, with increasing Fermi energy additional TSS carriers screen the potential fluctuations, and $\Gamma$ reduces with $1/\sqrt{|E_\mathrm{D}-E_\mathrm{F}|}$~\cite{Skinner2013a}.

When we compare these results with the literature on 3D-TI bulk crystals (see Table~\ref{tab:puddles}), we find reasonable consistencies in both $\Gamma$ and $\lambda$. The particularly large values of $\Gamma$ observed in this work and also in BiSbTeSe$_2$ \cite{Knispel2017} confirm the expectation that the Coulomb impurities resulting from compensation doping lead to large potential fluctuations. One can also see that even in simple binary compounds without compensation doping (Bi$_2$Se$_3$, Bi$_2$Te$_3$, and Sb$_2$Te$_3$), it is difficult to reduce the $\Gamma$ value to less than a few meV. It would be very useful if one could find a way to reduce the $\Gamma$ in a TI to $\sim$1 meV level, which is desirable for raising the critical temperature of the quantum anomalous Hall effect \cite{Chong2020,Lippertz2022} or for realizing stable Majorana bound states in proximitized TI nanowires~\cite{Heffels2023}.

Recent theory predicted \cite{Huang2021} that the large $\Gamma$ on the surface remains effectively unchanged when reducing the 3D bulk systems to quasi-2D thin films. Our result confirms this prediction. Furthermore, the disorder effect is found to be exacerbated in bulk-insulating films where the chemical potential lies around the middle of the bulk band gap. In these films, the $\Gamma$ value of around 40 meV is observed, and it is barely affected by the screening due to a superconductor at the opposite side of the 7 nm thick film. 
It is still to be seen if the large $\Gamma$ remains even in quasi-1D nanowires, but if it does not change much in nanowires, our result implies that the subband spacing needs to be larger than $\sim$10 meV to investigate the subband physics, given that $\Gamma$ in (Bi$_{1-x}$Sb$_x$)$_2$Te$_3$ is at least a few meV.  Since the subband spacing is given by $\Delta E \approx \hbar v_\mathrm{D} (2\pi/L)$ with $L$ is the perimeter length of the nanowire, for $v_\mathrm{D}=\qty{4e5}{\meter\per\second}$, the nanowire perimeter should be less than $\sim$160 nm. This is consistent with a recent work \cite{Muenning2021}, where the gate-voltage-dependent resistance oscillations due to subband crossings were observed in (Bi$_{1-x}$Sb$_x$)$_2$Te$_3$ nanowires with the diameter of $\sim$30 nm.

It is useful to note that there is a weak but positive correlation between $\Gamma$ and the FWHM of the Landau-level peaks, as shown in the inset of Fig. \ref{fig:EDvsX}(c). This can be attributed to the broadening due to a spatially more rapidly varying potential in the more disordered films:
Let us assume a quasi-classical motion of an electron in 9 T in a potential $\phi(\textbf{r})$ that changes smoothly with the length scale of $\lambda$. The electrons will make cyclotron orbits in a strip of the width $l_H(9\,T)\approx$ 8.5 nm while drifting along equipotential lines $e\phi(\textbf{r})$ = const. If the potential change is slow ($\lambda \gg l_H$), we expect a broadening of the Landau levels proportional to quantum smearing stemming from the drift motion~\cite{Champel2009}, which causes FWHM $\propto \l_H |\nabla_\textbf{r}e\phi(\textbf{r})|$. Therefore, the strength of the potential variation $|\nabla_\textbf{r}e\phi(\textbf{r})|$ is the relevant parameter that determines the measured Landau-level peak width. 
In our experiment, we made a coarse spatial sampling of the STM spectra (discussed in Sec.~\ref{sec:LandauLevelandEDP}) and performed the Landau level measurements at the location where the spectra is the sharpest. The above argument suggests that such a location corresponds to the local extrema of the fluctuating potential. When the  potential is more rapidly varying, $|\nabla_\textbf{r}e\phi(\textbf{r})|$ would be larger around such extrema and results in broader Landau level spectra. Since $\lambda$ is essentially independent of $\Gamma$ (see Table \ref{tab:overviewsamples}), $|\nabla_\textbf{r}e\phi(\textbf{r})|$ would be larger for larger $\Gamma$. 

In conclusion, we found that the effect of Coulomb disorder in (Bi$_{1-x}$Sb$_x$)$_2$Te$_3$ films is relatively strong, causing the amplitude of the potential fluctuations $\Gamma$ of at least a few meV. The $\Gamma$ gets worse as the films become more bulk-insulating and becomes as large as $\sim$40 meV. The best compromise is achieved at the Sb concentration $x \approx 0.65$, at which the films are quasi-bulk-insulating and $E_{\rm F}$ is within $\sim$10 meV from the Dirac point. This conclusion is consistent with the report by Scipioni \etal~\cite{Scipioni2018}. The length scale of the potential fluctuations $\lambda$ is found to be \JB{13 -- 54} nm, which gives a constraint on the device size if the potential fluctuation is detrimental to the physics to be studied. The (Bi$_{1-x}$Sb$_x$)$_2$Te$_3$ films should be primarily used for such applications where the chemical potential fluctuations are not detrimental but the bulk-insulating nature is crucial, such as spintronics \cite{Breunig2022, Dang2023}.

The raw data used in the generation of main and supplementary figures are available in Zenodo with the identifier 10.5281/zenodo.13889543.
.

\acknowledgments{We are grateful for insightful discussions with Achim Rosch, Thomas B\"omerich, Leonard Kaufhold and Thomas Lorenz.
This project has received funding from the European Research Council (ERC) under the European Union's Horizon 2020 research and innovation programme (grant agreement No 741121) and was also funded by the Deutsche Forschungsgemeinschaft (DFG, German Research Foundation) under CRC 1238 - 277146847 (Subprojects A04 and B06) as well as by the DFG under Germany's Excellence Strategy - Cluster of Excellence Matter and Light for Quantum Computing (ML4Q) EXC 2004/1 - 390534769.}

\appendix

\section{Quantum-well states}
\label{Sec:QWS}

\begin{figure}[h]
\includegraphics[scale=0.9]{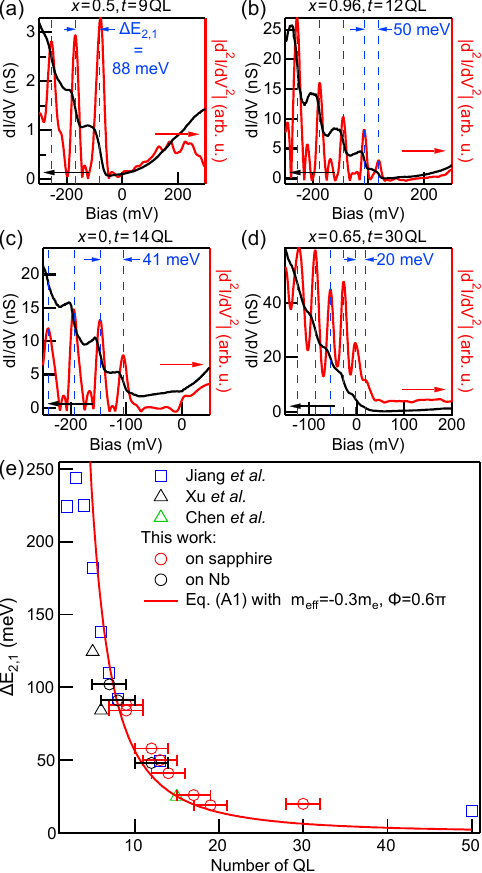}
\caption{\textbf{Quantum-well states in ultra-thin (Bi$_{1-x}$Sb$_x$)$_2$Te$_3$ thin films.} (a-d) \dIdU{} spectra (black trace) taken on samples with various $x$ and $t$ indicated for each panel. The steps in \dIdU{} spectra are attributed to subbands originating from the quantum confinement of the bulk valence band. The subband spacing is obtained as the distance between the peaks appearing in the numerically-calculated \ddIddU{} (red trace). (e) Thickness dependence of the subband spacing, taken as the energy difference between the first and second subbands, $\Delta E_{2,1}$. The data from the literature for $\mathrm{Bi}_2\mathrm{Te}_3${} by Chen {\it et al.} \cite{Chen2012} and by Xu {\it et al.} \cite{Xu2015}, as well as for $\mathrm{Sb}_2\mathrm{Te}_3${} by Jiang {\it et al.} \cite{Jiang2012, Jiang2012a},  are also plotted. The solid line is a fit to Eq.~\ref{eq:QWS}. Setpoints: (a,b) $I_0=0.2$~nA, $V_0=300$~mV; (a) $V_\mathrm{mod}=10$~mV$_\mathrm{p}$; (b) $V_\mathrm{mod}=5$~mV$_\mathrm{p}$; (c) $I_0=0.2$~nA, $V_0=50$~mV, $V_\mathrm{mod}=2$~mV$_\mathrm{p}$; (d) $I_0=-5$~nA, $V_0=-150$~mV, $V_\mathrm{mod}=3$~mV$_\mathrm{p}$.}
\label{fig:QWS}
\end{figure}

To analyze the expected energy levels of the quantum-well states formed by the quantum-confinement effect along the thickness direction, we utilize the phase accumulation model~\cite{Yang2009, Becker2010, Chen2021} and compare the result with the experimentally observed bulk-band splitting.
The quantization condition of the bulk band is based on the Bohr-Sommerfeld quantization rule
$$
2k(E)t+\phi=2\pi n,
$$
with $\phi=\phi_\mathrm{s}+\phi_\mathrm{v}$ the sum of phase shifts on reflection on the sapphire (or Nb) substrate and the vacuum barrier, and $k(E)$ is the energy-dependent wave vector of electrons propagating along the surface normal inside the film. Approximating the bulk valence band in (Bi$_{1-x}$Sb$_x$)$_2$Te$_3$ as quasi-free-electron-like, we get a spacing between the first two subbands for a film of thickness $t$ as
\begin{equation}\label{eq:QWS}
\Delta E_{2,1} = E_{\mathrm{v}_2}- E_{\mathrm{v}_1} = \frac{\hbar^2}{2m_\mathrm{eff}}\left(\frac{\pi}{t^2}\right)\cdot (3 \pi -\phi),\\
\end{equation}
with $m_\mathrm{eff}$ the effective mass of the bulk valence band which is approximated to be parabolic. 
We compare our experimental data with this simple model, along with the data for $\mathrm{Bi}_2\mathrm{Te}_3$ \cite{Xu2015, Chen2012} and for $\mathrm{Sb}_2\mathrm{Te}_3$ \cite{Jiang2012, Jiang2012a}, as shown in Fig.~\ref{fig:QWS}.

For $t$ of up to $\sim$20~QL, all experimental data follow Eq.~\ref{eq:QWS} with a reasonable effective mass $m_\mathrm{eff}\approx\qty{-0.3}{\m_e}$~\cite{Koehler1976,vonMiddendorff1973} and $\phi\approx 0.6\pi$, when we account for the experimental uncertainty ($\sim$2~QL) in the estimation of $t$. The significant deviations for films thicker than 25~QL are likely artificial, since the subband splitting $\Delta E_{2,1}$ is far below the experimentally observed band-edge broadening $\delta E \approx$ 30 -- 50 meV.

\section{Confirmation of rigid-band shift}
\label{sec:TechnicalDetailsDisorderPotentialCharacterization}

\begin{figure}[h]
\includegraphics[scale=1]{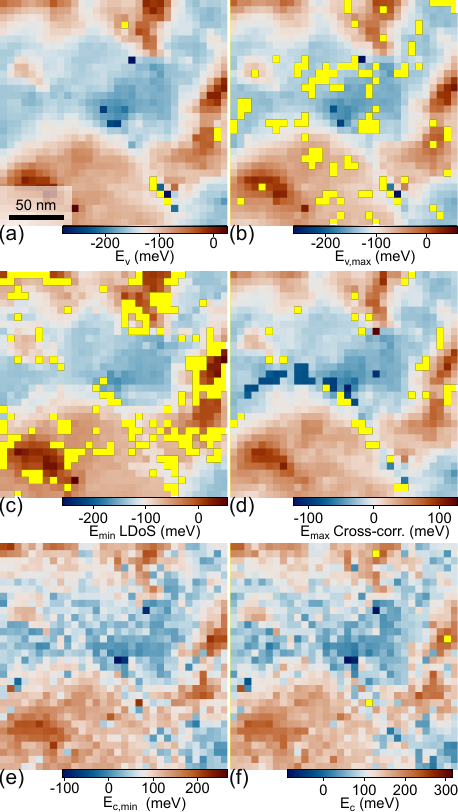}
\caption{\textbf{Illustration of the rigid-band shift.} (a-f) Maps of the same field of view of the spatial variations of $E_\mathrm{v}$, $E_\mathrm{v,max}$, minimum of LDoS , maximum of the cross-correlation, $E_\mathrm{c,min}$, and $E_\mathrm{c}$, as indicated in the panels. All maps show the same features, validating the underlying assumption of a rigid shift of the band structure due to local band bending caused by Coulomb disorder. Yellow pixels indicate failures in the numerical procedure and were disregarded. The potential map shown in Fig.~\ref{fig:Potential}(c) is computed as the arithmetic mean of all quantities shown here.}
\label{fig:ADisorder}
\end{figure}

Within the assumption of a rigid-band shift, one expects $e\phi=E_\mathrm{v}-\langle E_\mathrm{v} \rangle=E_\mathrm{c}-\langle E_\mathrm{c} \rangle=E_\mathrm{D}-\langle E_\mathrm{D} \rangle$. Hence, we ensured consistency of our procedure to calculate $e\phi$ by determining $E_\mathrm{v}$, $E_\mathrm{c}$, the variations of the minimum in the LDoS (which is the approximate $E_\mathrm{D}$) and the maximum in the cross-correlation between each spectrum and the average spectrum. Additionally, for computational convenience, we defined onset energies of the bulk valence band ($E_\mathrm{v,max}$) and the bulk conduction band ($E_\mathrm{c,min}$). As an example, we show the maps of all calculated quantities in Fig.~\ref{fig:ADisorder} highlighting that the spatial distribution is indeed the same for all of them. To mitigate isolated failures of the numerical determinations, we average over all maps of the six quantities mentioned above to compute the $e\phi$ in this paper.

\section{Analysis of the Landau level spectra}
\label{sec:LandauLevelDisorder}

\begin{figure}[t]
\includegraphics[scale=.9]{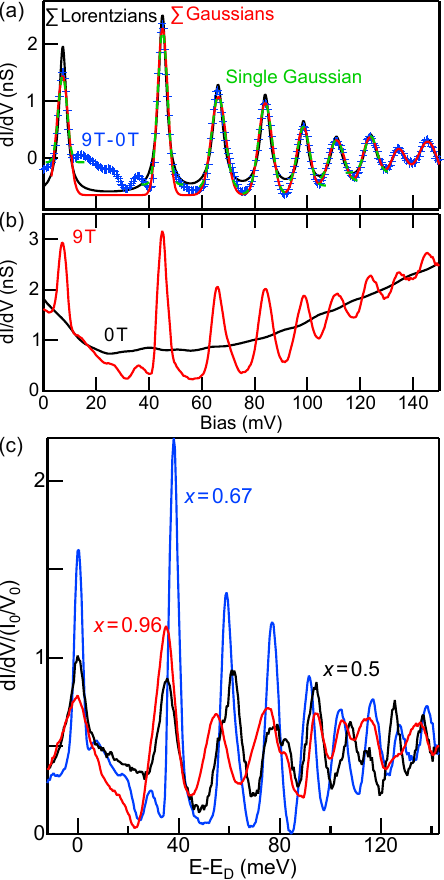}
\caption{\textbf{Background subtraction and fitting of the Landau level spectra.} (a) The \dIdU{} spectrum of the $x=0.67$ sample in 9 T after subtracting the 0 T data that were used as the background. The data before subtraction are the same as in Fig.~\ref{fig:LandauLevel}(b). Three different fits are shown: (i) sum of several Gaussians (red) to fit the whole spectrum, (ii) sum of several Lorentzians (black) to fit the whole spectrum, (iii) fitting only the vicinity (typically $\pm$10 meV) of each peak with a single Gaussian function (cyan). (b) Comparison of the raw spectra in 0 T and 9 T, showing that the 0 T data are a reasonable choice as the background. (c) Comparison of the 9 T spectra for different $x$ values shown in Fig.~\ref{fig:LandauLevel}(b,c,d) after background subtraction. The least disordered sample ($x=0.67$) clearly shows narrower Landau-level peaks than others.}
\label{fig:appendixLL}
\end{figure}

We characterized the width of the Landau levels shown in Fig.~\ref{fig:LandauLevel} and in supplement with the following procedure: First, the spectrum taken at 0 T is subtracted from the spectra in applied magnetic fields, to account for the background. Next, we fit the vicinity (typically $\pm$10 meV) of each Landau-level peak with a single Gaussian function to extract the full width at half maximum (FWHM). We find that differences in fitting all Landau levels with a sum of many Gaussians are negligible, and hence used the single peak fitting for computational simplicity for all FWHM values given in the paper.

Interestingly, the line-shape of the Landau levels is found to be more Gaussian than the frequently used Lorentzian~\cite{Jiang2012a, Hanaguri2010, Storz2018}. The details of the Landau level line-shape are a subtle issue due to instrumentation factors, but theoretically, a Gaussian line-shape with $\mathrm{FWHM} \propto \l_H |\nabla_\textbf{r}e\phi(\textbf{r})|$ is expected for quantum smearing from drift motion~\cite{Champel2009}. Examples of the background subtraction and fitting are shown in Fig.~\ref{fig:appendixLL}.


\section{Defect density}
\label{Sec:Defects}

\begin{figure}[h]
\includegraphics[scale=0.6]{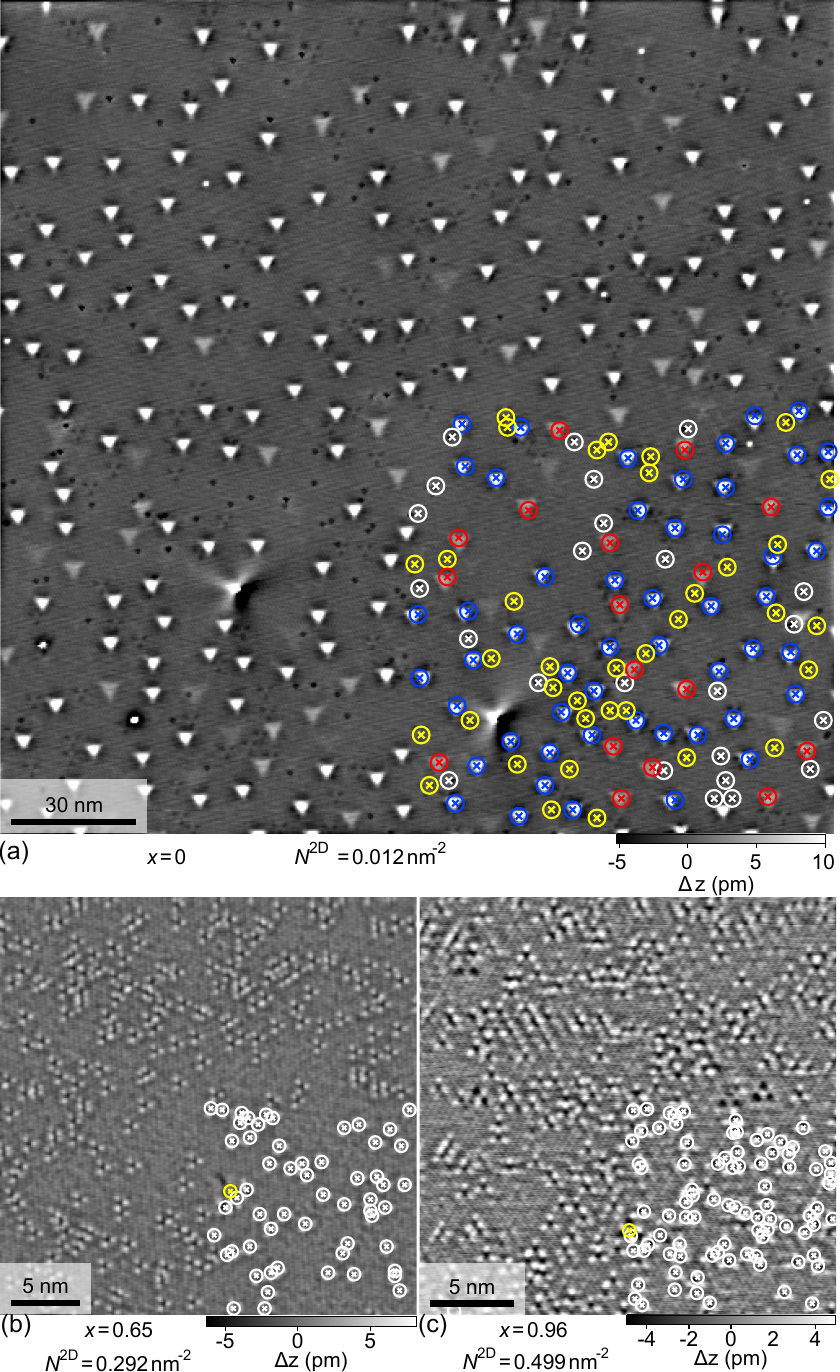}
\caption{\textbf{Atomic defects in (Bi$_{1-x}$Sb$_x$)$_2$Te$_3$ films.} (a,b,c) Typical topographs of samples with $x$ = 0, 0.65, and 0.96. The contrast in these images is enhanced by applying a standard high-pass filter to suppress long wavelength variations. Different types of defects are marked with crosses of different colors in the lower right quadrant of each image. See Ref.~\citenum{Lin2021} for the details of the types of defects. (a,b,c) The 2D defect density ($N^{\rm 2D}$) are 0.012, 0.292, and 0.499~nm$^{-2}$. Setpoints: (a,b,c) $V_0=900$~mV; (a) $I_0=0.1$~nA; (b) $I_0=5$~nA; (c) $I_0=0.2$~nA.}
\label{fig:Defects}
\end{figure}

It is well established~\cite{Dai2016,Lin2021} that STM can directly image the native defects near the surface, \textit{i.e.} in the top quintuple layer, of TIs. Counting the defects in the three typical STM images shown in Fig.~\ref{fig:Defects} gives the 2D defect densities ($N^{\rm 2D}$) indicated in the caption. The defect densities \num{1.2e19}, \num{2.9e20}, and \qty{5e20}{\per\cubic\cm} given in Table I are calculated as $N_\mathrm{d}$ = $N^{\rm 2D}$/1 nm for the $x$ = 0, 0.65, and 0.96 films.

\JB{
\section{Comparison of defect distributions and potential fluctuation}
\label{Mapping}}

\begin{figure}[h]
\includegraphics[scale=0.9]{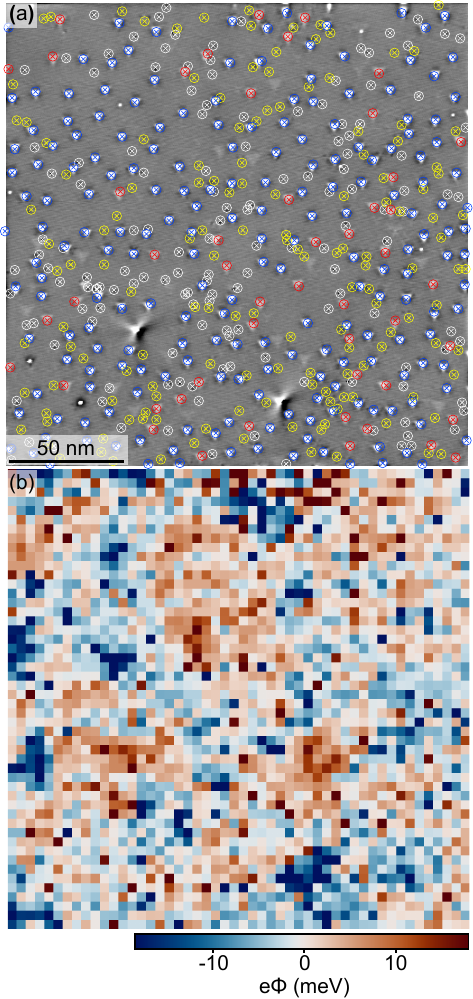}
\JB{\caption{Direct comparison of the defect distribution map (a) [same data as shown in Fig.~\ref{fig:Defects}(a)] and the potential map (b) for the same field of view measured on the sample MBE4 2022 Apr 07 with $x = 0$.}
\label{fig:Mapping}}
\end{figure}

\JB{Figure~\ref{fig:Mapping} shows a direct comparison of the defect distribution map and the potential map for the same field of view. One can see that there is no apparent correlation between them. This is understandable because the formation of charge puddles is dictated by a long-range statistical distribution of the charged acceptors and donors \cite{Skinner2013, Skinner2013a} and there is no reason that the short-range spatial distribution of the charged acceptors/donors has a clear correlation with the local potential. Note also that STM is only sensitive to the charged acceptors/donors near the surface, but the charge puddles will be dictated by the distribution of charges in the whole thickness.}

\newpage

\end{document}


\title{Supplemental Material for ``Characterizing the chemical potential disorder in the topological insulator (Bi$_{1-x}$Sb$_x$)$_2$Te$_3$ thin films''}
	
\maketitle
\vspace{-11mm}

\begin{center}
Jens Brede, Mahasweta Bagchi, Adrian Greichgauer, Anjana Uday, Andrea Bliesener, Gertjan~Lippertz, Roozbeh Yazdanpanah, Alexey Taskin, and Yoichi Ando\\
{\it Physics Institute II, University of Cologne, D-50937 K\"oln, Germany}
\end{center}

\vspace{3mm}
In this supplment, we give an overview of the data evaluated for the entries in Table I. 
\vspace{3mm}
\subsection{Sample MBE4 2022 Apr 28B ($x=0$)}
\vspace{-3mm}
In this Bi$_2$Te$_3$ film, which had similar properties as the other Bi$_2$Te$_3$ film discussed in the next section, the Landau-level spectroscopy was possible.

\begin{figure*}[h]
	\centering
	\includegraphics[scale=.77]{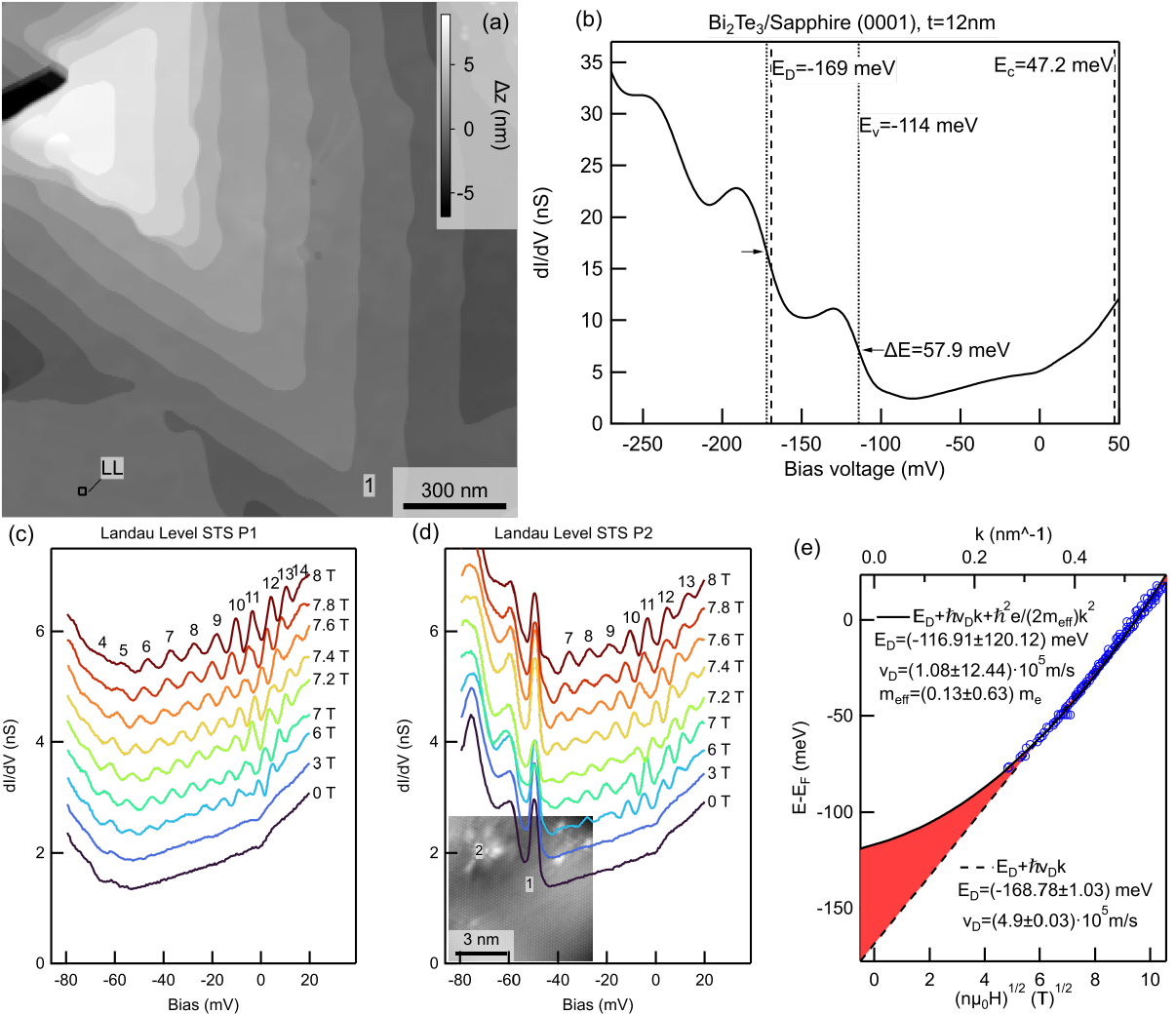}
	\caption{\linespread{1.05}\selectfont{}Sample MBE4 2022 Apr28B ($x=0$). 
	 (a) Topography including the locations `1' and `LL', where the large-scale STS (Point 1) and Landau-level STS (Point LL) were taken. (b) Large-scale STS with  E$_\mathrm{v}$, E$_\mathrm{c}$, and E$_\mathrm{D}$ marked. (c) Landau level STS taken at Point 1; the same data are also shown in Fig. 3(a) of the main text. (d) Landau level STS taken at P2, which is indicated in the inset of (d) and was on top of a defect giving rise to a defect state at $\sim -50$~meV, just above the bulk valence band top. (e) Dispersion relation and Dirac point energy are determined by fitting the Landau levels (c,d) as discussed in the article.}
	\label{fig:2022Apr28BMBE4_17nmBSTx0}
\end{figure*}

\newpage
\subsection{Sample MBE4 2022 Apr 07 ($x=0$)}

\begin{figure*}[h]
	\centering
	\includegraphics[scale=.65]{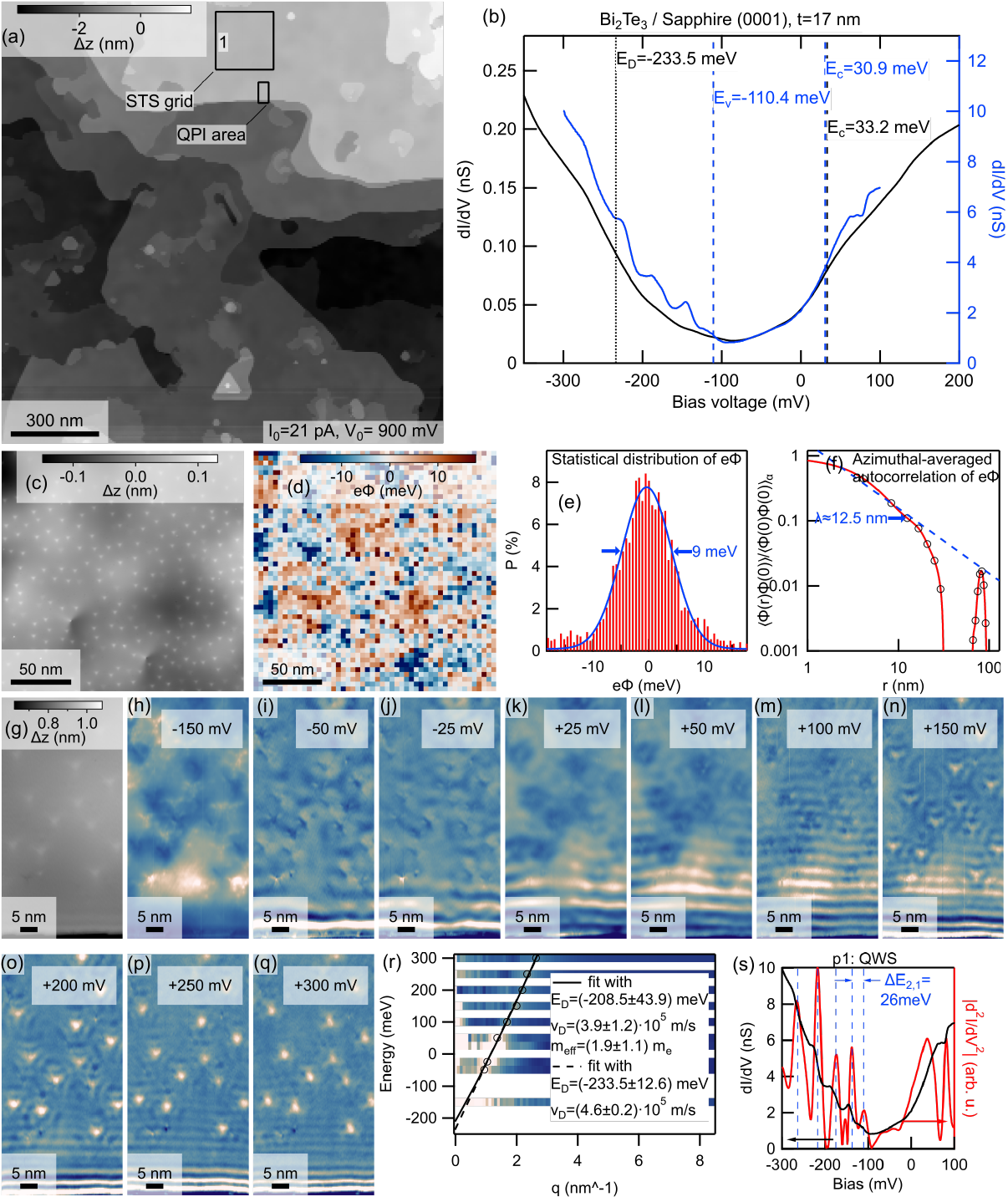}
	\caption{\linespread{1.05}\selectfont{}
Sample MBE4 2022 Apr07 ($x=0$). 
(a) Topography with the location of the STS grid and quasi-particle interference (QPI) measurements indicated. (b) Large scale STS with E$_\mathrm{v}$, E$_\mathrm{c}$, and E$_\mathrm{D}$ marked. The black trace corresponds to the average of the $dI(V-\phi)/dV$ curve of the STS grid and blue is the single high-resolution spectrum shown also in (s). (c) Topography of the STS grid region. (d,e) Spatial and statistical distribution of $e\phi$. (f) Azimuthal average of the 2D-autocorrelation of (d). (g) Topography of the QPI area. (h-q) Bias-dependent QPI. (r) the Fourier transform of (h-q). Fitting the data to the dispersion relation of the TSS for the full formula (solid line) as well as neglecting the $k^2$ term (dashed line) using the scaling relation $q=1.5k$. (s) STS taken at the position marked "1" in (a) with the quantum well states (QWS) indicated.}
	\label{fig:2022Apr07BMBE4_17nmBSTx0}
\end{figure*}

For this sample, the Dirac point energy was determined from the bias-dependent standing waves due to quasiparticle interference visible when the topological surface state (TSS) is scattered at a step-edge as depicted in Fig. \ref{fig:2022Apr07BMBE4_17nmBSTx0} (h-q). The energy-dependent scattering vector $q(E)$ is determined by calculating the Fourier transform of (h-q) which results in the dispersion relation $E(q)$ shown in (r). The scaling $q=1.5k$ relates $E(q)$ to the dispersion relation of the TSS which is given as
$E=E_\mathrm{D}+\hbar v_\mathrm{D}k+(\hbar^2e k^2)/(2m_\mathrm{eff})$.

\newpage
\subsection{Sample MBE3 2022 Oct28D2 ($x=0.5$)}
Here, extended data for the sample shown in Fig.~1 and Fig.~2 of the main text is given. 
\begin{figure*}[h]
	\centering
	\includegraphics[scale=.7]{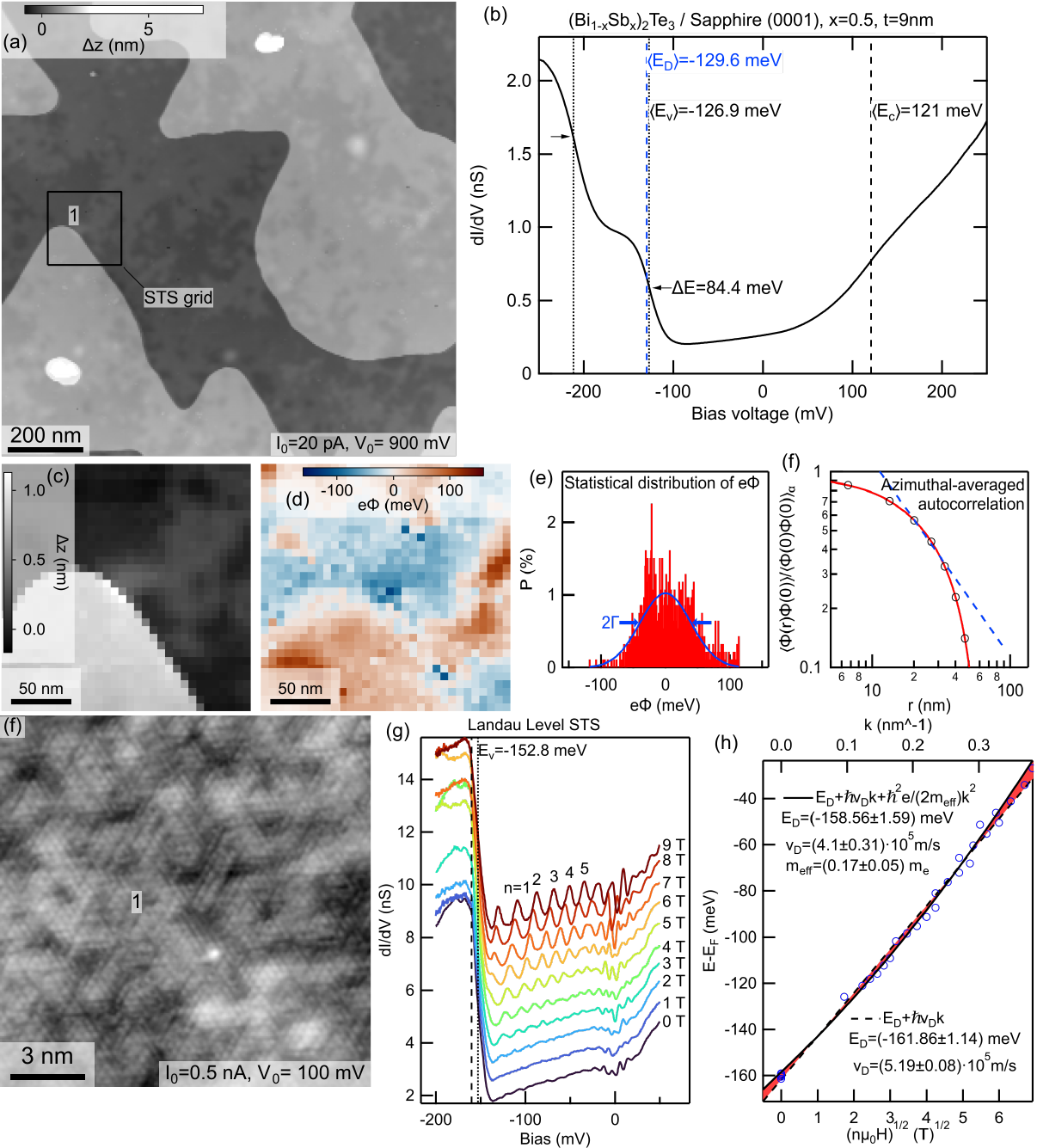}
	\caption{\linespread{1.05}\selectfont{}Sample MBE3 2022 Oct28D2 ($x=0.5$). (a) Topography (in parts also depicted in Fig.~1) with location of STS grid and Landau level STS (Point 1) indicated. (b) The average of the $dI(V-\phi)/dV$ curve of the STS grid with E$_\mathrm{v}$, E$_\mathrm{c}$, E$_\mathrm{D}$ and splitting of the bulk valence band QWS ($\Delta E$) marked. (c,d,e,f) Same data as Fig.~2~(b,c,d,e). (g) Atomic resolution of the same area where the Landau-level STS (shown in (h) was taken at position `1'. (i) The unassigned peak below $n=1$ also shifts with changing $\mu_0H$ similar to a previous report by Okade et al.~\cite{Okada2012}. (h) Dispersion relation and Dirac point energy are determined by fitting the Landau levels (g) as discussed in the article.}
	\label{fig:2022Oct28D2MBE3_10nmBSTx05}
\end{figure*}	

\newpage
\subsection{Sample MBE3 2022 Oct28D2 FC ($x=0.5$)}
\begin{figure*}[h]
	\centering
	\includegraphics[scale=.8]{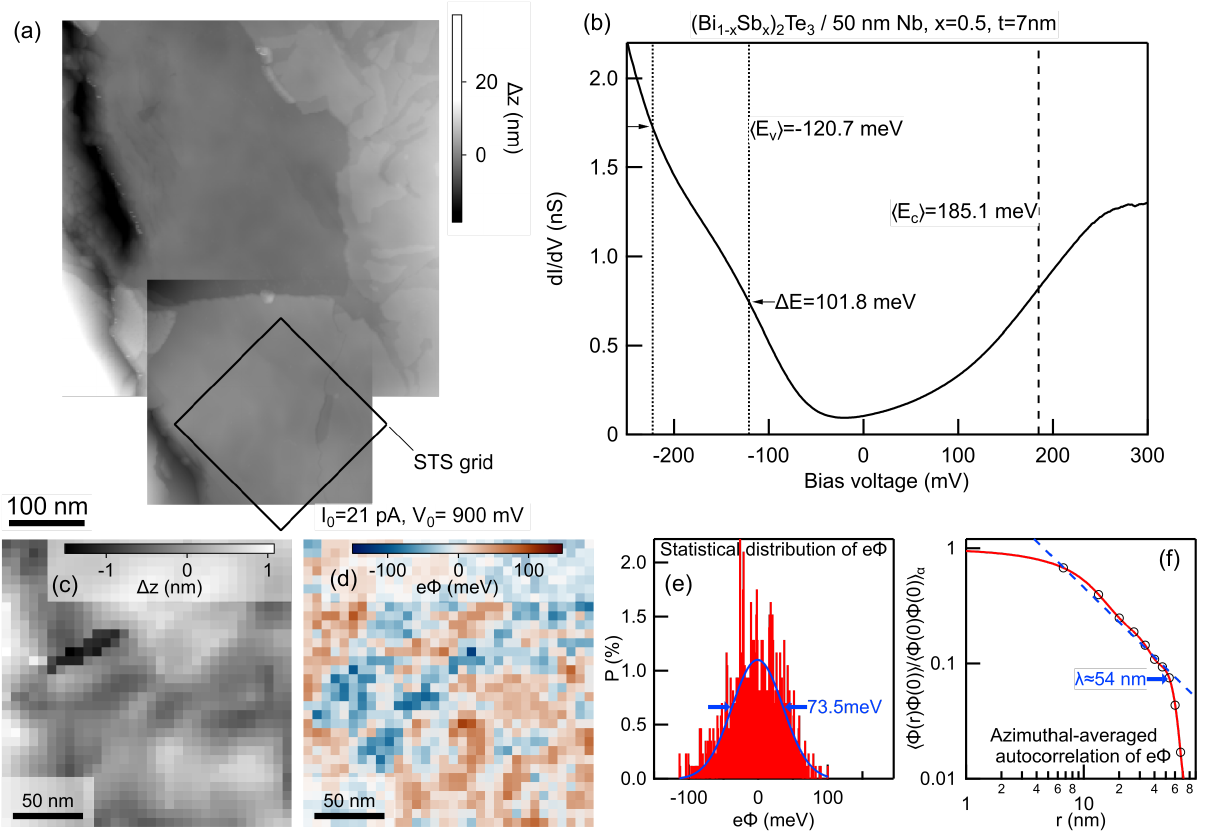}
	\caption{\linespread{1.05}\selectfont{}Sample 2022 Oct28D2 MBE3 FC ($x=0.5$). (a) Several topographies were stitched together to illustrate the position where the STS grid was taken. (b) The average of the $dI(V-\phi)/dV$ curve of the STS grid with E$_\mathrm{v}$, E$_\mathrm{c}$, and splitting of the bulk valence band QWS ($\Delta E$) marked. (c) Topography of the STS grid region. (d,e) Spatial and statistical distribution of $e\phi$. (f) The azimuthal-averaged autocorrelation of the spatial distribution of $e\phi$.}
	\label{fig:2022Oct28D2MBE3_10nmBSTx05_50nmNb_FC}
\end{figure*}

After the measurements of the BST film on sapphire, it was used to prepare a flip-chip (FC) sample with $\sim 50$~nm Nb and measured again, with the results summarized in Fig.~\ref{fig:2022Oct28D2MBE3_10nmBSTx05_50nmNb_FC}. Interestingly, no significant changes in the disorder potential are seen when comparing Figs.~\ref{fig:2022Oct28D2MBE3_10nmBSTx05} and \ref{fig:2022Oct28D2MBE3_10nmBSTx05_50nmNb_FC}.

\newpage
\subsection{Sample MBE3 2022 Oct28C2 ($x=0.5$)}

We tested the reproducibility by analyzing a sample grown with nominally identical conditions as sample MBE3 2022 Oct28D2 discussed in the previous section. Qualitatively, the data are consistent with the expected bulk-insulating behavior, but the sample was less $n$-type.

\begin{figure*}[htbp]
	\centering
	\includegraphics[scale=.61]{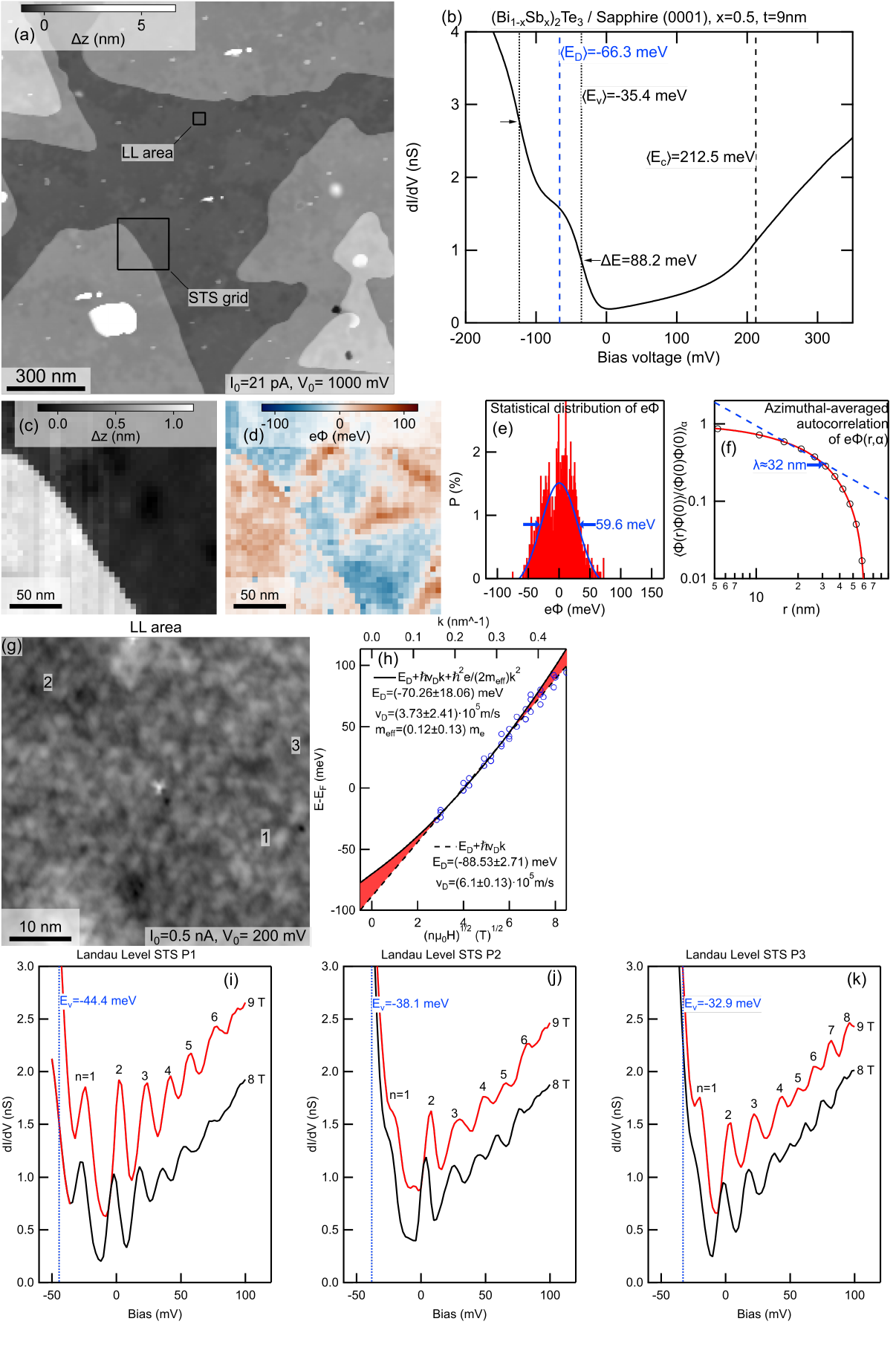}
	\caption{\linespread{1.05}\selectfont{}Sample MBE3 2022 Oct28C2 ($x=0.5$). (a) Topography with locations of STS grid and Landau level (LL) STS indicated.  (b) The average of the $dI(V-\phi)/dV$ curve of the STS grid with the average values of E$_\mathrm{v}$, E$_\mathrm{c}$, and splitting of the bulk valence band QWS ($\Delta E$) marked. (c) Topography of the STS grid region. (d,e) Spatial and statistical distribution of $e\phi$. (f) The azimuthal-averaged autocorrelation of the spatial distribution of $e\phi({\bf r})$. (g) Topography of LL area with positions 1-3 of the spectra shown in (i-k) indicated. (h) Dispersion relation and Dirac point energy are determined by fitting the Landau levels (i-k) as discussed in the article.}
	\label{fig:2022Oct28C2MBE3_10nmBSTx05}
\end{figure*}	

\newpage
\subsection{Sample MBE3 2022 Jun6C FC ($x=0.5$)}

\begin{figure*}[ht]
	\centering
	\includegraphics[scale=.85]{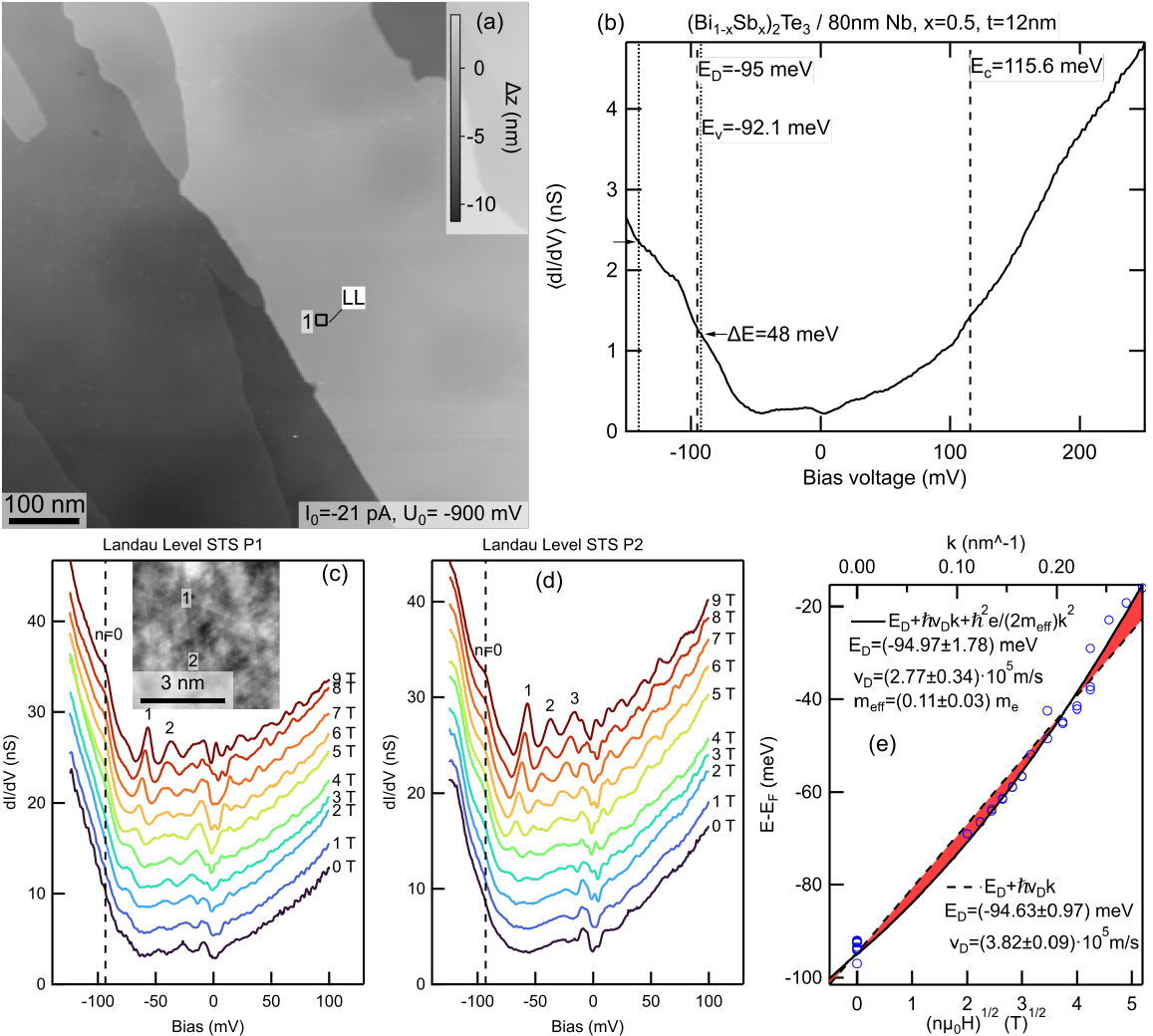}
	\caption{\linespread{1.05}\selectfont{}Sample MBE3 2022 Jun6C FC ($x=0.5$). (a) Topography with the locations of Landau level (LL) and large scale (`1') STS indicated. (b) The large scale STS with the values of E$_\mathrm{v}$, E$_\mathrm{c}$, E$_\mathrm{D}$, and splitting of the bulk valence band QWS ($\Delta E$) marked. (c,d) LL STS. Data in (d) is the same as in Fig.~3~(d). The atomic resolution of the LL area is shown in the inset of (c) with locations `1', `2' corresponding to spectra in (c,d). (e) Dispersion relation and Dirac point energy are determined by fitting the Landau levels (c,d) as discussed in the article.}
	\label{fig:2022Jun6CMBE3_FC_BSTx50_80nmNb}
\end{figure*}

This film had a nominal thickness of 19~nm, but the terrace of the film evaluated and shown in Fig.~\ref{fig:2022Jun6CMBE3_FC_BSTx50_80nmNb} below was determined to have a local thickness of 12~nm from the QWS spacing  ($\Delta E$).

\newpage
\subsection{Sample MBE2 2020 Oct 09B ($x=0.65$)}

\begin{figure*}[ht]
	\centering
	\includegraphics[scale=.7]{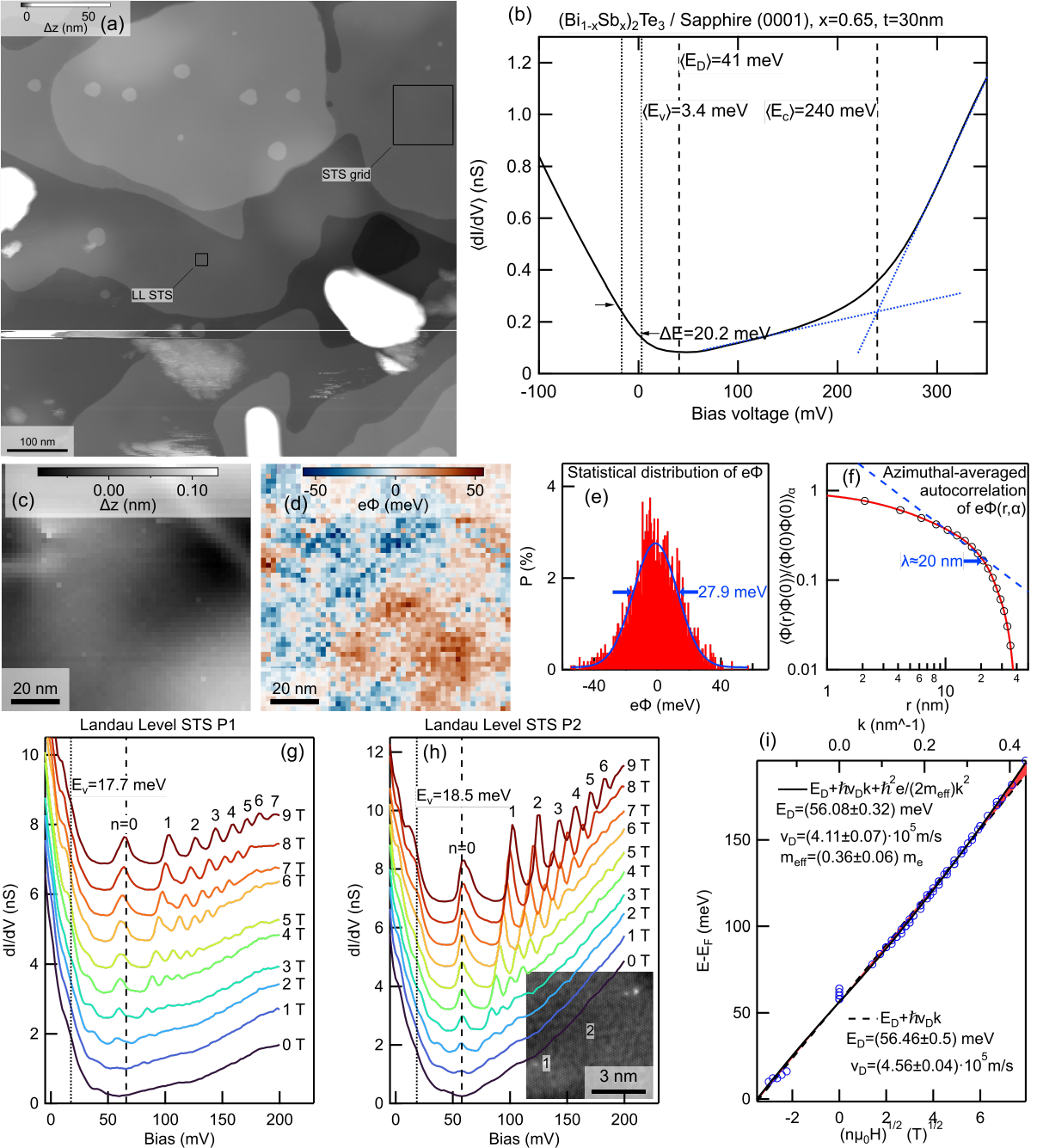}
	\caption{\linespread{1.05}\selectfont{}Sample MBE2 2020 Oct 09B ($x=0.65$). (a) Topography with locations of STS grid and Landau level (LL) STS indicated. (b) The average of the $dI(V-\phi)/dV$ curve of the STS grid with the average values of E$_\mathrm{v}$, E$_\mathrm{c}$, and E$_\mathrm{D}$ marked. The indicated splitting of the bulk valence band QWS ($\Delta E$) was determined from the higher resolution LL spectra. Since no conduction band QWS was resolved, we estimated $\langle \mathrm{E}_\mathrm{c}\rangle$ by linear extrapolation as indicated by the blue dotted lines. (c) Topography of the STS grid region. (d,e) Spatial and statistical distribution of $e\phi$. (f) The azimuthal-averaged autocorrelation of the spatial distribution of $e\phi({\bf r})$. The atomic resolution of the LL area is shown in the inset of (h) with indicated positions `1', `2'  where LL spectra shown in (g,h) were taken. (e) Dispersion relation and Dirac point energy are determined by fitting the Landau levels (g,h) as discussed in the article.}
	\label{fig:2020Oct9MBE2BSTx65}
\end{figure*}	

\newpage
\subsection{Sample MBE3 2022 Jun 05B2 ($x=0.67$)}
\begin{figure*}[ht]
	\centering
	\includegraphics[scale=.7]{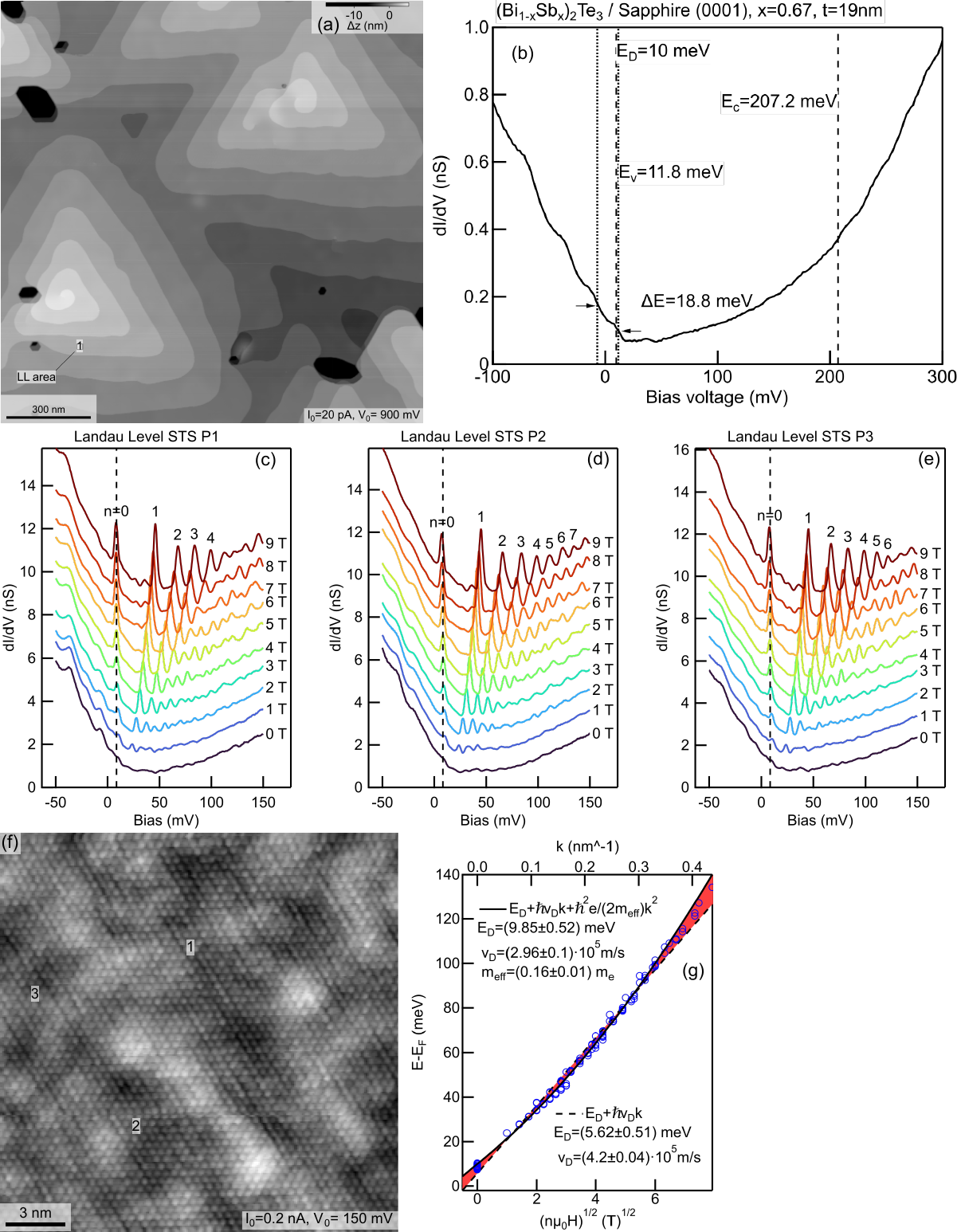}
	\caption{\linespread{1.05}\selectfont{}Sample MBE3 2020 Jun 05B2 ($x=0.67$). (a) Topography with the location of the Landau level (LL) and large scale (`1') STS indicated. (b) The large scale STS with the values of E$_\mathrm{v}$, E$_\mathrm{c}$, E$_\mathrm{D}$, and splitting of the bulk valence band QWS ($\Delta E$) marked. (f) Atomic resolution of the LL area with positions 1-3 where the Landau level STS (c,d,e) was recorded. Spectra in (d) are the same data as in Fig.~3~(b). (g) Dispersion relation and Dirac point energy are determined by fitting the Landau levels (c-e) as discussed in the article.}
	\label{fig:2022_June5B2_MBE3}
\end{figure*}	

\newpage
\subsection{Sample MBE1 2021 Sep 08 FC ($x=0.86$)}

\begin{figure*}[ht]
	\centering
	\includegraphics[scale=.7]{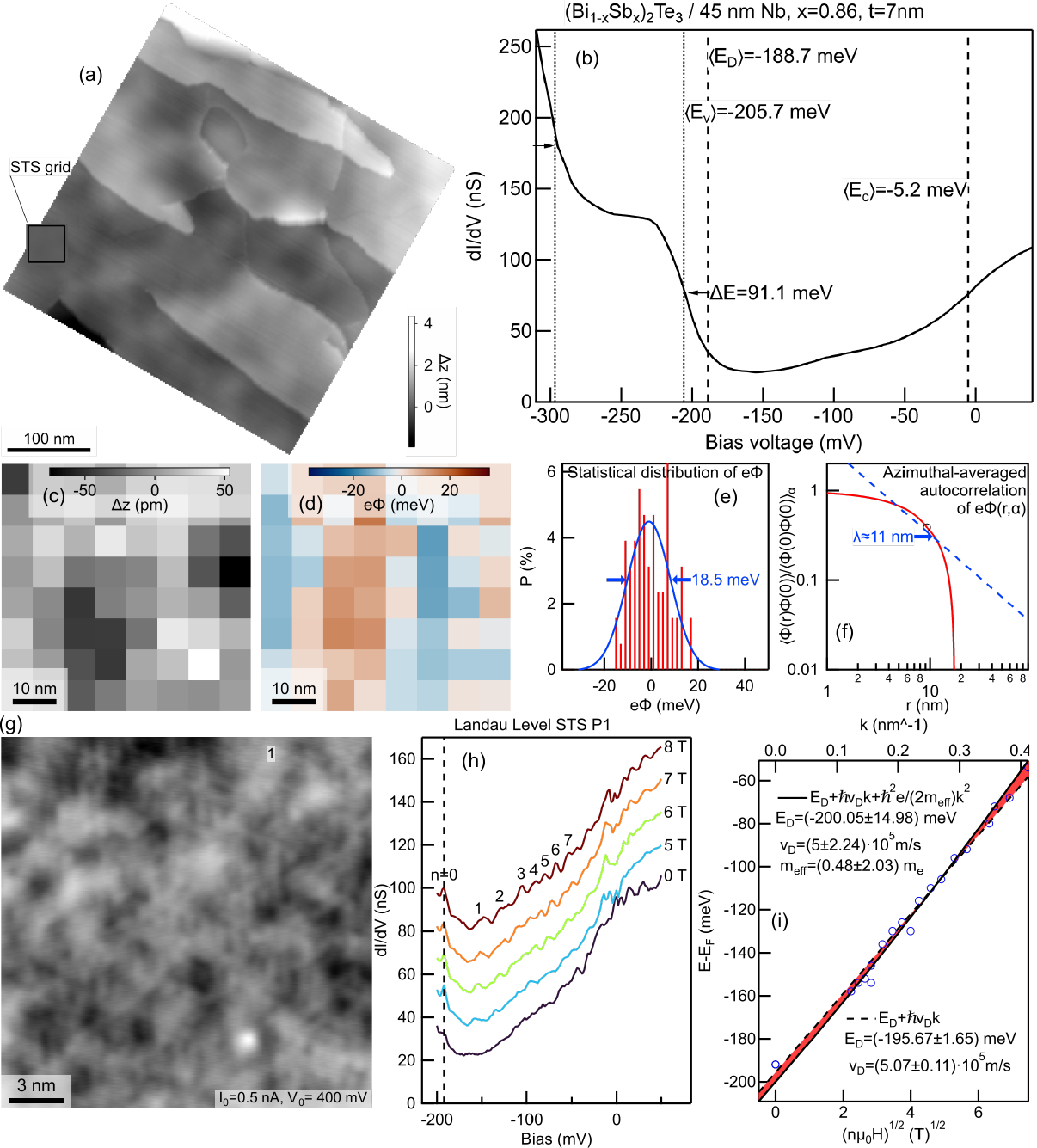}
	\caption{\linespread{1.05}\selectfont{}Sample MBE1 2021 Sep 08 FC (x=0.86). (a) Topography with locations of STS grid indicated. (b) The average of the $dI(V-\phi)/dV$ curve of the STS grid with the average values of E$_\mathrm{v}$, E$_\mathrm{c}$, and splitting of the bulk valence band QWS ($\Delta E$) marked. (c) Topography of the STS grid region. (d,e) Spatial and statistical distribution of $e\phi$. (f) The azimuthal-averaged autocorrelation of the spatial distribution of $e\phi({\bf r})$. (g) High-resolution topography STS grid region where also the Landau level (LL) spectra were taken at position `1'. (h) LL spectra. (i) Dispersion relation and Dirac point energy are determined by fitting the Landau levels (h) as discussed in the article.}
	\label{fig:2021Aug25MBE1_FC_8nmBSTx0}
\end{figure*}

This film had a nominal thickness of 17~nm, but the terrace of the film evaluated and shown in Fig.~\ref{fig:2021Aug25MBE1_FC_8nmBSTx0} was determined to have a local thickness of 7~nm from the QWS spacing  ($\Delta E$).

\newpage
\subsection{Sample MBE3 2022 Apr 13A ($x=0.96$)}
\begin{figure*}[ht]
	\centering
	\includegraphics[scale=.55]{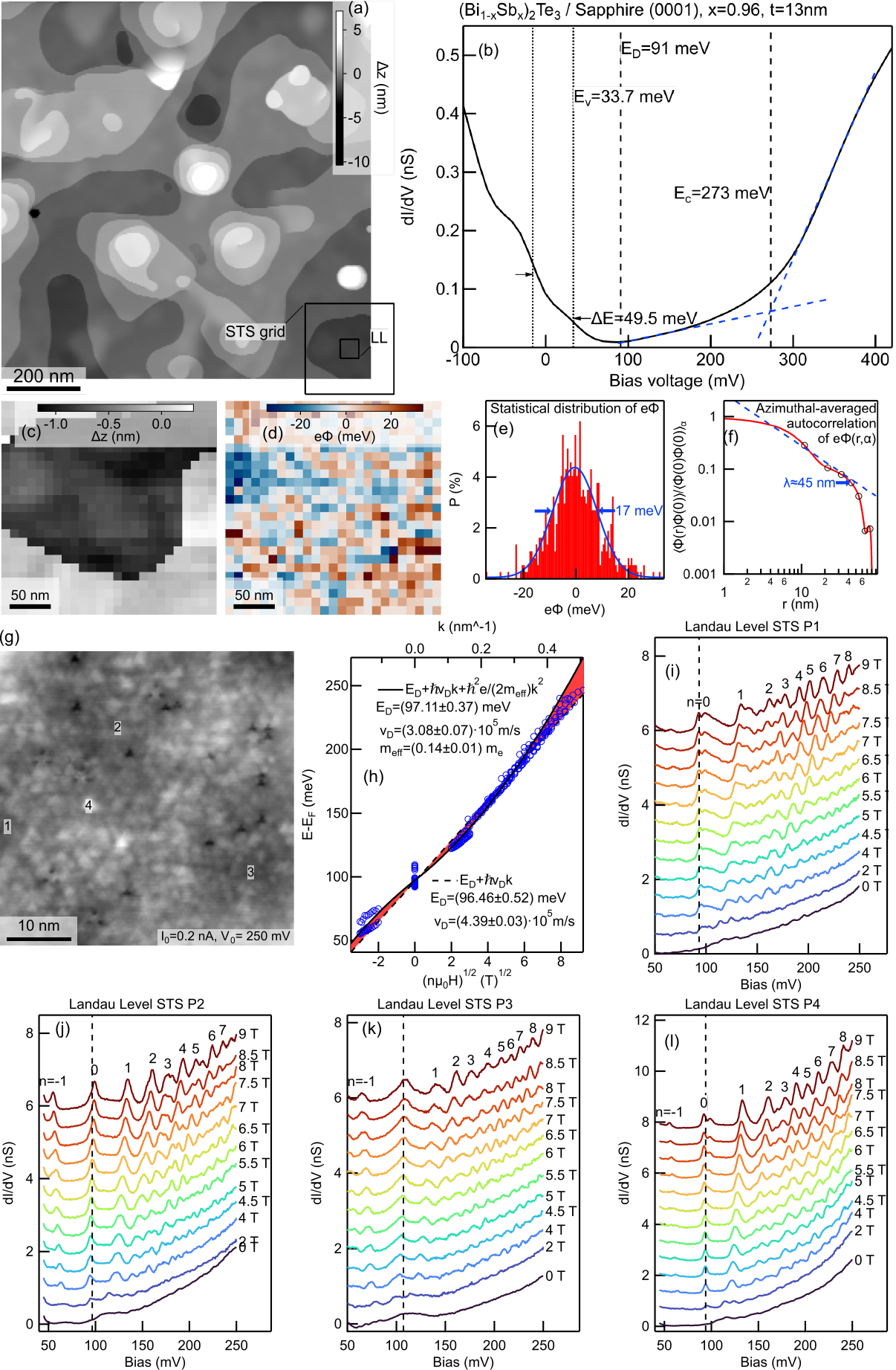}
	\caption{\linespread{1.05}\selectfont{}Sample MBE3 2022 Apr 13A ($x=0.96$). (a) Topography with locations of STS grid and Landau level (LL) area indicated. (b) The average of the $dI(V-\phi)/dV$ curve of the STS grid with the average values of E$_\mathrm{v}$, E$_\mathrm{c}$, E$_\mathrm{D}$, and splitting of the bulk valence band QWS ($\Delta E$) marked. (c) Topography of the STS grid region. (d,e) Spatial and statistical distribution of $e\phi$. (f) The azimuthal-averaged autocorrelation of the spatial distribution of $e\phi({\bf r})$. (g) Topography of the LL area with positions 1-4 where Landau level STS shown in (i-l) were taken. (h) Dispersion relation and Dirac point energy are determined by fitting the Landau levels shown in (i-l) as discussed in the article. The data shown in (j) are the same as in Fig.~3(c).}
	\label{fig:2022MBE3Apr13A_BSTx096}
\end{figure*}